\documentclass[aps,prx,reprint,superscriptaddress,floatfix,amsmath,amssymb]{revtex4-2}
\usepackage[tworuled,vlined]{algorithm2e}
\usepackage{booktabs}
\usepackage{color}
\usepackage{enumitem}
\usepackage{graphicx}
\usepackage{hyperref}
\usepackage{physics}

\hypersetup{colorlinks=true,urlcolor=blue,citecolor=blue,linkcolor=blue,breaklinks=true}

\newcommand\ee{\mathrm{e}}%
\newcommand\ii{\mathrm{i}}%
\newcommand\bigO{\mathcal{O}}%

\newcommand\mbcrea[1]{\mathrm{#1}^\dagger}
\newcommand\mbannh[1]{\mathrm{#1}^{\mathchoice{\vphantom\dagger}{}{}{}}}
\newcommand\cee{\mbannh{c}}%
\newcommand\cdag{\mbcrea{c}}%
\newcommand\eff{\mbannh{f}}%
\newcommand\fdag{\mbcrea{f}}%

\newcommand\iv{\ii\nu}%
\newcommand\iw{\ii\omega}%
\newcommand\wmax{\bgroup\omega_{\mathrm{max}}\egroup}%

\newcommand\WW{\mathcal{W}}
\newcommand\FF{\mathrm{F}}
\newcommand\BB{\mathrm{B}}

\newcommand\Fbar{\mathrm{\bar F}}
\newcommand\Bbar{\mathrm{\bar B}}
\newcommand\Btilde{\mathrm{\bar{\bar{B}}}}

\newcommand\Norb{N_\mathrm{orb}}
\newcommand\Nbath{N_\mathrm{bath}}

\newcommand\linethrough[1]{%
    \rule[.2\baselineskip]{2em}{.5pt}%
    \hspace{-2em}%
    \makebox[2em]{#1}%
    }

\definecolor{purple}{rgb}{0.6, 0.2, 0.8}
\definecolor{orange}{rgb}{0.91, 0.41, 0.17}

\begin{document}

\hyphenation{par-ti-cle--hole Hart-ree--Fock}

\allowdisplaybreaks
\title{
Solving the Bethe--Salpeter equation with exponential convergence}
\author{Markus Wallerberger}
\thanks{M. Wallerberger and H. Shinaoka contributed equally to this work.}
\affiliation{Institute of Solid State Physics, TU Wien, 1040 Vienna, Austria}

\author{Hiroshi Shinaoka}
\thanks{M. Wallerberger and H. Shinaoka contributed equally to this work.}
\affiliation{Department of Physics, Saitama University, Saitama 338-8570, Japan}

\author{Anna Kauch}
\affiliation{Institute of Solid State Physics, TU Wien, 1040 Vienna, Austria}

\begin{abstract}
The Bethe--Salpeter equation plays a crucial role in understanding the physics 
of correlated fermions, relating to optical excitations in solids as well as 
resonances in high-energy physics.  Yet, it is notoriously difficult to control 
numerically, typically requiring an effort that scales polynomially with energy 
scales and accuracy.  This puts many interesting systems out of computational 
reach.

Using the intermediate representation and sparse modeling for two-particle 
objects on the Matsubara axis, we develop an algorithm that solves the Bethe--Salpeter
equation in $\bigO(L^8)$ time with $\bigO(L^4)$ memory, where $L$ grows only 
logarithmically with inverse temperature, bandwidth, and desired accuracy.
This opens the door for computations in hitherto inaccessible regimes.
We benchmark the method on the Hubbard atom and on the multi-orbital weak-
coupling limit, where we observe the expected exponential convergence to the 
analytical results.  We then showcase the method for a realistic impurity 
problem.
\end{abstract}
\maketitle

% =============================================
\section{Introduction}
% =============================================

The rapid increase in available computational resources has propelled material calculations into a new era.
Computational methods based on the density functional theory are widely used for material design.
One of the remaining grand challenges is computing dynamical response determining the functionalities of  materials, especially those with strong electronic correlations.  In theory,
excitons, magnons, and other composite excitations can only be described by two-particle correlation functions (optical conductivity, susceptibilities).  At the core of calculating these lie equations at the two-particle level, in particular the Bethe--Salpeter equation~\cite{Bethe_Salpeter,GellMann_Low}.

\begin{figure}
    \centering
    \begin{tabular}{lllll}
\toprule
\textsf{\bfseries\footnotesize (a)} & Linear data size & Truncation & Runtime & Memory  \\ 
       &           & error      & scaling & scaling \\
\midrule
Sparse & $L=\bigO(\log(\beta\wmax))$ & $\bigO(\ee^{-\alpha L})$  & $\Norb^6 L^8$ & $\Norb^4 L^4$ \\ 
Dense  & $N=\bigO(\beta\wmax)$       & $\bigO(N^{-1-\gamma})$      & $\Norb^6 N^4$ & $\Norb^4 N^3$ \\
\bottomrule
      \end{tabular}
\\[0.5em]
\includegraphics[width=\columnwidth]{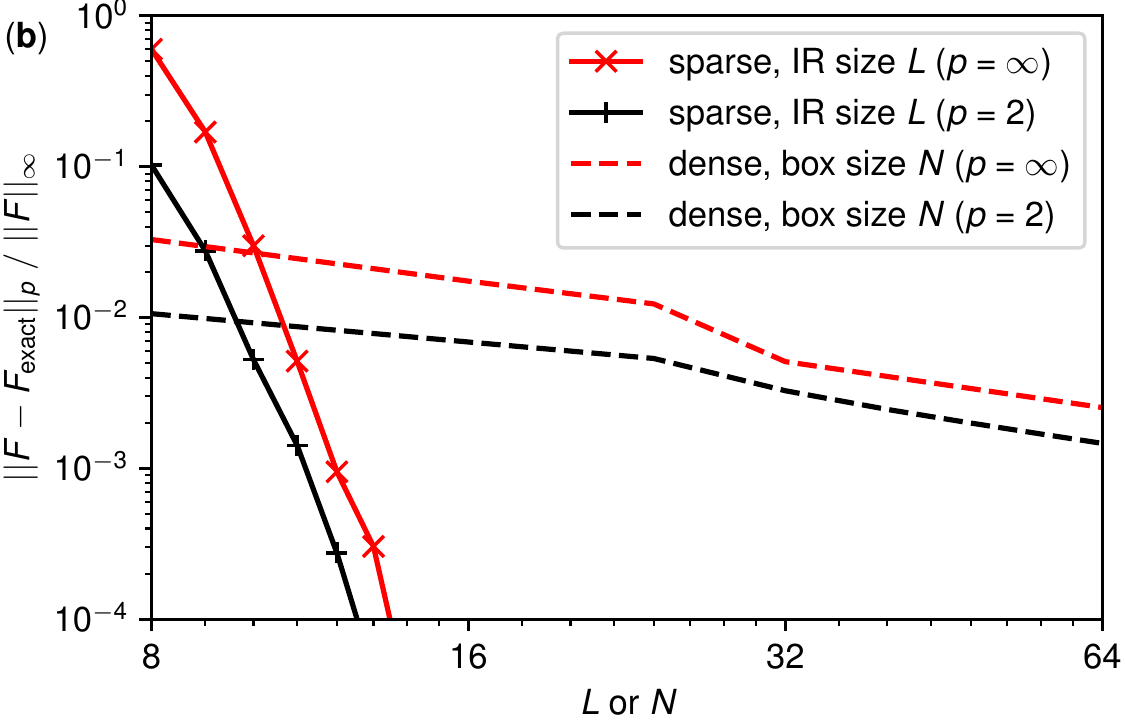}
    \caption{
Solving the BSE using sparse modeling with a cut\-off $L$ (this work) compared with using a dense frequency box
of linear size $N$: (a) nominal truncation error and runtime/memory scaling,
where $\Norb$ is the number of orbitals, $\beta\wmax$ is the bandwidth in units of temperature, and $\gamma$ is the power in inverse frequencies to which the asymptotics is known analytically; (b) actual scaling of the error for the Hubbard atom (cf. Sec.~\ref{sec:atom}), where black and red lines are the average ($p=2$) and
maximum ($p=\infty$) deviation from the exact result, respectively.
    }
    \label{fig:teaser}
    \vspace{-0.5cm}
\end{figure}

The Bethe--Salpeter equation (BSE) is a basic computational tool in a multitude of methods: from {\it ab initio} approaches~\cite{Blase2020,OnidaRMP,Aryasetiawan_1998,Takada2001,Maggio2017,Pavlyukh2020} through many-body methods in condensed matter~\cite{GeorgesRMP,Kunes11,Rohringer2012, Hafermann2014,Lin2012,Otsuki2014,RohringerRMP,Metzner2012,dmf2rg,Kugler2018, DeDominicis,Vasiliev98,Bickers04} but also high-energy~\cite{bse_in_qcd, bse_in_qcd2} and nuclear physics~\cite{nuclear-physics-textbook,parquet_nuclear}.
It relates the sum $F$ of all two-particle scattering channels to a smaller irreducible set of diagrams, $\Gamma$:
\begin{equation}
    \vcenter{\hbox{\includegraphics[scale=.68]{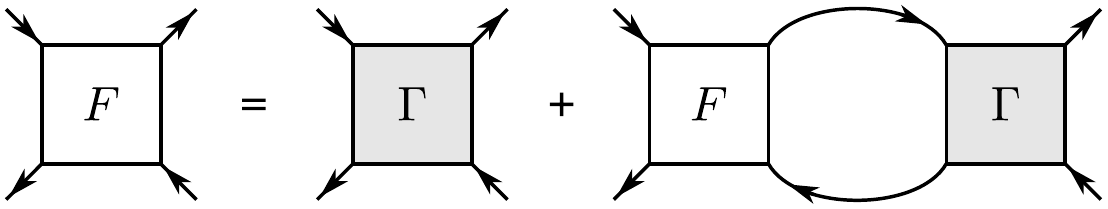}}},
    \label{eq:bse-gfx}
\end{equation}
where $F$ and $\Gamma$ depend on three frequencies and four orbital indices~\footnote{The number of orbital indices (by which we mean spin, orbital and for lattice systems also momentum indices here) can often be reduced by using symmetry relations. The number of momentum indices can be further reduced to three using momentum conservation.  Eq.~\eqref{eq:bse-gfx} is in the particle--hole channel. For details see Sec.~\ref{sec:bse} and Ref.~\onlinecite{RohringerRMP}.} (cf. Sec.~\ref{sec:bse} for details).

Solving the BSE is notoriously difficult since it scales unfavorably in the number of orbitals and frequencies.
To reduce the computational effort most {\it ab initio} calculations at the two-particle level are performed with the static approximation, where only the zero frequency component is taken~\cite{Blase2020,OnidaRMP,Bickers1997,Vilk1997,Valenti2019,fRG_ab_initio}, or with a reduced frequency dependence~\cite{Takada2001, Maggio2017, Pavlyukh2020}. The fully dynamical calculations~\cite{Kunes11, Lin2012, Rohringer2012, Hafermann2014,Otsuki2014, Galler2017,Kauch2020}
are quickly stopped by the curse of dimensionality at only few orbitals or high temperatures.
This renders many interesting regimes such as low-temperature phases of materials inaccessible.

Fully dynamical calculations mostly use a dense box of Matsubara frequencies.  Solving the BSE then
requires $\bigO(N^3)$ memory and, even worse, $\bigO(N^4)$ computational time, where $N$ is the number
of frequencies along each side of the box (see Fig.~\ref{fig:teaser}).  The na\"ive truncation error decays
as $\bigO(1/N)$.   These exponents are problematic since $N$ is 
proportional to the bandwidth, $\wmax$, and to the inverse temperature, $\beta=1/T$, with a large prefactor
due to substantial finite-size effects of the box.  Chemical accuracy is then usually far out of reach
and error control becomes difficult.

In this paper we offer a substantial step towards numerically low-cost evaluation of the BSE. Based on the previously developed concept of sparse modeling of the two-particle correlation functions~\cite{Shinaoka:2018cg,Shinaoka:2020ji}, we directly solve the BSE in the sparse representation (intermediate representation, IR~\cite{shinaoka2017compressing}). As illustrated in Fig.~\ref{fig:teaser}, the  polynomial scaling of the maximal (red curves) and average (black curves) relative error with the linear system size (linear frequency box size $N$) is replaced by faster than exponential decay with the size of the IR representation $L$. Moreover, $L$ grows only logarithmically with $\beta\wmax$~\cite{Chikano:2018gd}. With this method we achieve not simply an improvement in scaling, but a paradigm shift: from brute force high performance computing to data compression and systematic error control. The concept of sparse modeling is already widely adopted for {\it ab initio} calculations at the one-particle level~\cite{Li:2020eu,Wang2020,NomotoPRL2020,NomotoPRB2020,NomuraRR2020,Isakov2020,Witt2021}. Here, we pave the way to taking dynamical {\it ab initio} calculations with sparse modeling to the two-particle level.   

In order to keep the paper self-contained and introduce the necessary concepts and notation, we review sparse modeling  in Sec.~\ref{sec:sparse}. In Sec.~\ref{sec:dia-eqns}, we will then show that these ideas carry over to diagrammatic
equations such as the Bethe--Salpeter equation, which we rewrite in the compressed form. In Sec.~\ref{sec:convolution}, we then develop a method to also compress
all intermediate results (summation over inner propagators) needed in solving the
BSE (sparse convolution). The final step needed to obtain the solution is to solve the resulting least-squares problem, which we present in Sec.~\ref{sec:expansion}.  We then benchmark our method on two challenging systems: in Sec.~\ref{sec:bench}, we compare
sparse modeling to the analytic results for the Hubbard atom as well as for the weak-coupling limit
of a multi-orbital impurity.  In Sec.~\ref{sec:impurity}, we then show the results for  a realistic impurity problem.
We conclude with a summary and outlook in Sec.~\ref{sec:conclusion}.

% =============================================
\section{Review: Sparse modeling}
\label{sec:sparse}
% ============================================
We review sparse modeling before its use in solving the BSE.
The concept of sparse modeling is based on a generic feature of imaginary frequency data:
in transitioning numerical data from real to imaginary (Matsubara) frequencies, we lose information as features of
correlation functions get smeared out.
That information loss makes it difficult
to reconstruct the original real-frequency signal, but at the same time allows imaginary frequency
data to be compressed extremely efficiently~\cite{shinaoka2017compressing,Chikano:2018gd,OtsukiSparseAnaCont}.

The essence of this approach is twofold: (i)~the truncated intermediate representation (IR) provides an efficient basis for representing 
Matsubara Green's functions and (ii)~the basis coefficients can be inferred from Matsubara data at special sampling frequencies 
in a quick and stable fashion.
In the following, we review the use of IR in representing one-particle and two-particle response functions, focusing specifically on the Green's function.
For an extended review on the IR, we refer 
the reader to Refs.~\onlinecite{shinaoka2021efficient} and \onlinecite{Otsuki:2020fn}.

% ---------------------------
\subsection{One-particle response functions}
\label{sec:sparse2}
% ---------------------------

A one-particle response function can be written as ~\cite{Mahan}:
\begin{equation}
\hat{G}^\alpha(\iw)=\int_0^\beta\dd{\tau}\ee^{\iw\tau}\langle T_{\tau}A^\alpha(\tau)B^\alpha(0)\rangle,\label{eq:giwdef}
\end{equation}
where $\tau$ denotes imaginary
(Euclidean) time, $T_{\tau}$ indicates time ordering, $\alpha\in\{\mathrm{B},\mathrm{F}\}$
denotes fermionic or bosonic statistics, $A$ and $B$ are operators,
and $\iw$ is a fermionic (bosonic) Matsubara frequency for $\alpha=\mathrm{F}$
($\alpha=\mathrm{B}$).
Equation~(\ref{eq:giwdef}) and the associated
spectral function $\rho(\omega)$ in real frequency $\omega$ are
connected through analytic continuation~\cite{Mahan}:
\begin{equation}
\hat{G}^\alpha(\iw)=\int_{-\wmax}^\wmax\dd{\omega'}
\underbrace{
\frac{\omega'^{\delta_{\alpha,\BB}}}{\iw-\omega'}
}_{K^\alpha(\iw,\omega')}\rho^\alpha(\omega'),\label{eq:friedholm}
\end{equation}
where we require the spectral function to be of the form $\rho(\omega)=\sum_i\rho_i\delta(\omega-\omega_i)$
with $\omega_i\in[-\wmax,\wmax].$ (Note that the bosonic kernel has been regularized using an additional factor $\omega'$, consistent with analytic continuation practice.)

The kernel in a Fredholm integral equation of the first kind, $K^\alpha$ admits a singular value expansion \cite{Hansen:SVE}:
\begin{equation}
K^\alpha(\iw,\omega')=\sum_{l=0}^\infty\hat{U}_l^\alpha(\iw)S_l^\alpha V_l^{\alpha}(\omega'),\label{eq:sve}
\end{equation}
where  $S_l^\alpha$ are the singular values in strictly decreasing order, $S_0 > S_1 > \ldots > 0$,
$\hat{U}_l^\alpha$ are the left singular functions, which form an orthonormal set
on the Matsubara frequencies, and $V_l^\alpha$ are the right singular functions, which form an orthonormal set
on the real frequencies~%
\footnote{
In general, the complex conjugate of $V_l^\alpha(\omega')$ appears in Eq.~(\ref{eq:sve}).
For convenience, we choose $V_l^\alpha(\omega')$ as real functions.
In this gauge, the left singular functions, when transformed to $\tau$, are real because the Fredholm kernel is real in $\tau$}.
The singular functions $\hat{U}_l^\alpha$ and $V_l^\alpha$ are the so-called IR basis functions~\cite{shinaoka2017compressing}.

We can use the left singular functions of the kernel as
representation for the Green's function~\cite{shinaoka2017compressing}:
\begin{equation}
\hat G^\alpha(\iw)=\sum_{l=0}^{L-1} \hat{U}_l^\alpha(\iw)G^\alpha_l+\epsilon_L,\label{eq:gtaurep}
\end{equation}
where $G^\alpha_l = S_l \sum_i \rho_i V^{\alpha}_l(\omega_i)$ is a basis coefficient and $\epsilon_L$ is an error term associated
with truncating the series. 
One can prove that $S_l$ drop faster than any power with $l$%
~\cite{singvalsuper}, and there is numerical evidence for 
faster-than-exponential decay~\cite{Chikano:2018gd}.
One empirically finds logarithmic growth of singular values with respect
to increasing the energy cutoff~\cite{Chikano:2018gd}, $S_L/S_0=\bigO(\log(\beta\wmax))$,
and also that the right singular functions are bounded, i.e., there
exists a $V_\mathrm{max}$ such that $|V_l(\omega)|<V_\mathrm{max}$. This implies that the
truncated representation (\ref{eq:gtaurep}) converges faster than exponentially, $\epsilon_L=\bigO(S_L)$,
and is substantially more compact than a polynomial representation.

In order to efficiently extract the basis coefficients $G_l$ from
$\hat{G}(\iw)$, we exploit the fact that the roots 
of the singular functions are similar in structure to the roots of orthogonal polynomials~\cite{Rokhlin:1996sve}.
Hence, we choose a set of sampling frequencies close to the sign changes of $\hat{U}_L(\iw)$:
\begin{equation}
\WW^\alpha=\{\iw_1^\alpha,\iw_2^\alpha,\ldots,\iw_L^\alpha\},\label{eq:wsample2}
\end{equation}
and turn Eq.~(\ref{eq:gtaurep}) into an ordinary least squares fit~\cite{Li:2020eu}:
\begin{equation}
    G^\alpha_l = \arg\min_{G_l} \sum_{\iw\in\WW^\alpha} \bigg| \hat G^\alpha(\iw) - \sum_{l=0}^{L-1} \hat U_l^\alpha(\iw)G^\alpha_l \bigg|^2.
\end{equation}
An example of sampling frequencies is shown in Fig.~\ref{fig:self-energy}.

One empirically observes that the $|\WW|\times L$ fitting matrix $U_{nl}=\hat{U}_l^\alpha(\iw^\alpha_n)$ is well-conditioned provided that $\WW$ was chosen as in Eq.~(\ref{eq:wsample2})~\cite{Li:2020eu}.
This in turn implies the fitting error is consistent with the overall truncation
error $\epsilon_L$. Similar rules can be found for imaginary time.

% ---------------------------
\subsection{Two-particle response functions}
\label{sec:sparse4}
% ---------------------------

We now turn to the two-particle Green's function:
\begin{equation}
\begin{split} & \hat{G}(\iv_1,\iv_2,\iv_3,\iv_4)=
\int_0^\beta\dd{\tau_1}\dd{\tau_2}\dd{\tau_3}\dd{\tau_4}\\
& \quad\times\ee^{\iv_1\tau_1-\iv_2\tau_2+\iv_3\tau_3-\iv_4\tau_4}\langle T_{\tau}A(\tau_1)B(\tau_2)C(\tau_3)D(\tau_4)\rangle,
\end{split}
\label{eq:g4def}
\end{equation}
where $A,\ldots,D$ are now fermionic operators and, consequently,
$\iv_1$ to $\iv_4$ are fermionic Matsubara frequencies.

\begin{table}
    \centering
\begin{tabular}{rlll}
\toprule 
& $\iv$ & $\iv'$ & $\iw$\tabularnewline
\midrule
$T_{1} (\ldots)=$ & $\delta(\iv,+\iv_1)$ & $\delta(\iv',+\iv_4) $ & $\delta(\iw,\iv_1 - \iv_2)$ \tabularnewline
$T_{2} (\ldots)=$ & $\delta(\iv,+\iv_1)$ & $\delta(\iv',-\iv_3) $ & $\delta(\iw,\iv_1 - \iv_2)$ \tabularnewline
$T_{3} (\ldots)=$ & $\delta(\iv,+\iv_1)$ & $\delta(\iv',+\iv_4) $ & $\delta(\iw,\iv_1 + \iv_3)$ \tabularnewline
$T_{4} (\ldots)=$ & $\delta(\iv,+\iv_1)$ & $\delta(\iv',+\iv_2) $ & $\delta(\iw,\iv_1 + \iv_3)$ \tabularnewline
$T_{5} (\ldots)=$ & $\delta(\iv,+\iv_1)$ & $\delta(\iv',-\iv_3) $ & $\delta(\iw,\iv_1 - \iv_4)$ \tabularnewline
$T_{6} (\ldots)=$ & $\delta(\iv,+\iv_1)$ & $\delta(\iv',+\iv_2) $ & $\delta(\iw,\iv_1 - \iv_4)$ \tabularnewline
$T_{7} (\ldots)=$ & $\delta(\iv,-\iv_2)$ & $\delta(\iv',+\iv_4) $ & $\delta(\iw,\iv_2 + \iv_1)$ \tabularnewline
$T_{8} (\ldots)=$ & $\delta(\iv,-\iv_2)$ & $\delta(\iv',-\iv_3) $ & $\delta(\iw,\iv_2 + \iv_1)$ \tabularnewline
$T_{9} (\ldots)=$ & $\delta(\iv,-\iv_2)$ & $\delta(\iv',+\iv_4) $ & $\delta(\iw,\iv_2 + \iv_3)$ \tabularnewline
$T_{10}(\ldots)=$ & $\delta(\iv,-\iv_2)$ & $\delta(\iv',-\iv_3) $ & $\delta(\iw,\iv_2 - \iv_4)$ \tabularnewline
$T_{11}(\ldots)=$ & $\delta(\iv,+\iv_3)$ & $\delta(\iv',+\iv_4) $ & $\delta(\iw,\iv_3 + \iv_1)$ \tabularnewline
$T_{12}(\ldots)=$ & $\delta(\iv,+\iv_3)$ & $\delta(\iv',+\iv_4) $ & $\delta(\iw,\iv_3 - \iv_2)$ \tabularnewline
\bottomrule
\end{tabular}
\caption{Our choice of frequency translation tensor $T_r$ in Eq.~(\ref{eq:g4lehmann}), 
    with representations $r=1,\ldots,12$ corresponding to
    representations $5,\ldots,16$ of Ref.~\onlinecite{Shinaoka:2020ji} (representations 1 to 4 can be absorbed into the
    others by means of a partial fraction decomposition).
}
\label{tab:transl}
\end{table}

The Lehmann representation of Eq.~(\ref{eq:g4def}) can be cast
in the following form~\cite{Shinaoka:2018cg}:
\begin{equation}
\begin{split} 
& \hat{G}(\iv_1,\iv_2,\iv_3,\iv_4)=\sum_{r=1}^{12} \sum_{\nu\nu'\omega} T_r(\iv_1,\ldots,\iv_4; \iv, \iv', \iw) \\
& \ \times\int \dd{^3\omega}
K^\FF\!(\iv,\omega_1) K^\FF\!(\iv',\omega_2) K^\Bbar\!(\iw,\omega_3)
\rho_r(\omega_1,\omega_2,\omega_3),\\
\end{split}
\label{eq:g4lehmann}
\end{equation}
where $\iv, \iv'$ are fermionic Matsubara frequencies, $\iw$
is a bosonic Matsubara frequency, $\omega_1,\omega_2,\omega_3$
are real frequencies, and $T_r$ is a frequency translation tensor
defined in Table~\ref{tab:transl}.  This translation is necessary
in order to have a spectral function of the form:
\begin{equation}
\rho_r(\omega_1,\omega_2,\omega_3) =
   \sum_{ijk} A^{(r)}_{ijk} \delta(\omega_1-\omega_i) \delta(\omega_2-\omega_j) \delta(\omega_3-\omega_k),
   \label{eq:rho4}
\end{equation}
because the two-particle Green's function cannot be made compact
in any single frequency convention due to permutations induced by time ordering
in Eq.~(\ref{eq:g4def}).  Therefore, we require the sum over $r=1,\ldots,12$ different
representations in Eq.~(\ref{eq:g4lehmann}), unlike Eq.~(\ref{eq:friedholm}), 
where only one representation is sufficient.

Entering Eq.~(\ref{eq:g4lehmann}) is a product
of the one-particle kernels $K^\FF$ and $K^\BB$ from Eq.~(\ref{eq:friedholm}).
However, $K^\BB$ must be augmented in order to ensure proper decay of the expansion coefficients:
\begin{equation}
    K^\Bbar(\iw,\omega') = \frac{\omega'}{\iw-\omega'} + S^\Bbar_{0} \delta_{\iw,0} + \frac{S^\Bbar_{1} (1-\delta_{\iw,0})}{\iw},
    \label{eq:KBbar}
\end{equation}
where $S^\Bbar_{0}, S^\Bbar_{1}$ are arbitrary prefactors (we shift the remaining singular values by
$S^\BB_l \to S^\Bbar_{l+2}$).  A $\delta$-function at $\iw=0$ is not
included in the unaugmented kernel (\ref{eq:friedholm}) as it cannot be resolved by it; however, terms like those
are indeed present in the two-particle Green's function.

Since the one-particle kernels can be truncated, Eq.~(\ref{eq:g4lehmann}) implies that there also exists a truncated, compact representation for the two-particle Green's function analogous to Eq.~(\ref{eq:gtaurep}):
\begin{equation}
\begin{split} 
&\hat{G}(\iv_1,\ldots,\iv_4)=\sum_{r=1}^{12} \sum_{\nu\nu'\omega} T_r(\iv_1,\ldots,\iv_4; \iv, \iv', \iw) \\
&\qquad\times\sum_{l,l',m=0}^{L-1} \hat U_l^\FF(\iv) \hat U_{l'}^\FF(\iv') \hat U_m^\Bbar(\iw) G_{r,ll'm} + \epsilon_L,\\
\end{split}
\label{eq:g4ir}
\end{equation}
where $\hat U_l^\FF(\iv)$ is defined in Eq.~(\ref{eq:sve}) and $\hat U_l^\Bbar$ are the singular functions of the
bosonic kernel, augmented by the additional contributions in Eq.~(\ref{eq:KBbar}).
Since $G_{r,ll'm}$ is then given by projection of Eq.~(\ref{eq:rho4}):
\begin{equation}
    G_{r,ll'm} = S^\FF_l S^\FF_{l'} S^\Bbar_m
    \sum_{ijk} A^{(r)}_{ijk} V^{\FF}_l(\omega_i) V^{\FF}_{l'}(\omega_j) V^{\Bbar}_m(\omega_k),
    \label{eq:rho4ir}
\end{equation}
one again has faster than exponential convergence in Eq.~(\ref{eq:g4ir}), $\epsilon_L = \bigO(S_L)$.
Storing $G$ in the IR basis thus only requires $12L^3$ numbers, where $L$ is $\bigO(\log(\beta\wmax\epsilon^{-1}))$ and $\epsilon$ is the desired accuracy.

Extracting $G_{r,ll'm}$ from $G$ requires us to construct a fitting problem.  We choose the set of
sampling frequencies as:
\begin{equation}
\begin{split} 
\WW = \bigcup_{r=1}^{12} \big\{& (\iv_1,\ldots,\iv_4)
    : (\iv,\iv',\iw)\in \WW^\FF\times \WW^\FF\times \WW^\Bbar \\
& \mathrm{where}\ T_r(\iv_1,\ldots,\iv_4;\iv,\iv',\iw) \ne 0 \big\},
\end{split}
\label{eq:wsample4}
\end{equation}
i.\,e., the outer product of the one-particle sampling frequencies $\WW^\alpha$ from Eq.~(\ref{eq:wsample2})
according to the one-particle kernels in Eq.~(\ref{eq:g4lehmann}), transformed from the ``native'' frequencies
$(\iv, \iv, \iw)$ of each representation to the all-fermionic convention $(\iv_1,\ldots,\iv_4)$.
We show an example of sampling frequencies in Fig.~\ref{fig:quadrature}(a).

This choice ensures that the sampling points required to construct each $r,ll'm$ are present in $\WW$.
However, since the IR basis is overcomplete, every representation projects to the same set of 
frequencies and thus the ordinary least squares problem:
\begin{equation}
\begin{split}
   \min_G\!\Big[&\sum_{(\iv_i)\in\WW} \Big| \hat G(\iv_1,\ldots,\iv_4)
   -\!\sum_{r,ll'm}\!E^{\nu_1\nu_2\nu_3\nu_4}_{r,ll'm} G_{r,ll'm}\Big|^2\\
   &+ \lambda^2 \sum_{r,ll'm} |\Gamma^\mathrm{Tikh.}_{rll'm}G_{r,ll'm}|^2\Big],\\
\end{split}
   \label{eq:g4fit}
\end{equation}
where $E$ is the transformation tensor from the IR to the sampling frequencies from Eq.~(\ref{eq:g4ir}),
is ill-conditioned for $\lambda=0$.
A way to mitigate this problem is by using Tikhonov regularization, where one adds a term
to the cost function penalizing large fitting parameters $G_{r,ll'm}$ ($\lambda \ne 0$):
one, e.g., can use the {\em a priori} knowledge of the decay of basis coefficients 
from Eq.~(\ref{eq:rho4ir}) and 
enforce this by choosing $(\Gamma^{\mathrm{Tikh.}}_{rll'm})^{-1}=S^\FF_lS^\FF_{l'}S^\Bbar_m$ (for a more detailed discussion see Sec.~IV of Ref.~\onlinecite{Shinaoka:2018cg}).

As for the number of sampling frequencies, one finds $L^3 \le |\WW| \le 12L^3$, as sampling frequencies coming from multiple representations may coincide, and one typically has $|\WW|\approx 8 L^3$.

% =============================================
\section{Diagrammatic equations}
\label{sec:dia-eqns}
% =============================================

We will now discuss how to extend sparse modeling in order to solve two-particle
diagrammatic equations. Diagrammatic equations are an algebraic way to sum up whole classes
of diagrams, usually by invoking a topological argument.  They are the bread and butter of
the diagrammatic technique and at the heart of renormalization methods and embedding
techniques. As in Sec.~\ref{sec:sparse}, we will start with a one-particle example (the Dyson equation) and then go to the two-particle case.

% -----------------------------------
\subsection{Self-energy and vertex basis}
\label{sec:dyson}
% -----------------------------------

For simplicity, we again start with the one-particle case, where we will introduce the
tools that we later use for the two-particle case.

\begin{figure}
    \centering
    \includegraphics[width=\columnwidth]{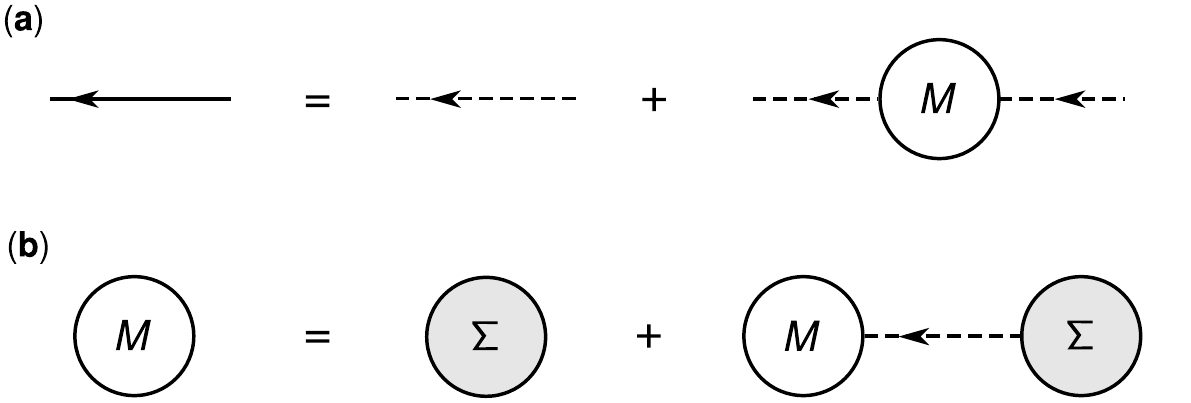}
    \caption{One-particle diagrammatic equations: (a) relation of the full Green's 
    function $G$ (solid line), non-interacting Green's function $G_0$ (dashed line) and
    full one-particle vertex $M$; (b) Dyson equation in the vertex form, where $\Sigma$
    is the self-energy (irreducible one-particle vertex).
    }
    \label{fig:dyson}
\end{figure}

Let $\hat G_0$ be
the non-interacting one-particle Green's function, i.e., Eq.~(\ref{eq:giwdef}) with $\langle\ldots\rangle$
replaced by the average $\langle\ldots\rangle_0$ over the non-interacting system. It is
related to the full Green's function via
\begin{equation}
    \hat G(\iv) = \hat G_0(\iv) + \hat G_0(\iv)\,\hat M(\iv)\,\hat G_0(\iv),
    \label{eq:M}
\end{equation}
where $\hat M$ is the full one-particle vertex encoding all interactions in the system,
and $\iv$ is the fermionic Matsubara frequency~%
\footnote{$\hat M$ is usually known as scattering matrix or improper self-energy~\cite{fetter-walecka}.}.
(We restrict ourselves to fermionic systems for simplicity.) Diagrammatically, Eq.~(\ref{eq:M}) is
shown in Fig.~\ref{fig:dyson}(a).  The full vertex $\hat M$ is related to its 
irreducible counterpart, the self-energy $\hat\Sigma$, via the Dyson equation:
\begin{equation}
    \hat M(\iv) = \hat \Sigma(\iv) + \hat M(\iv)\,\hat G_0(\iv)\,\hat\Sigma(\iv),
    \label{eq:dyson}
\end{equation}
which is shown diagrammatically in Fig.~\ref{fig:dyson}(b).

\begin{figure}
    \centering
    \includegraphics[width=\columnwidth]{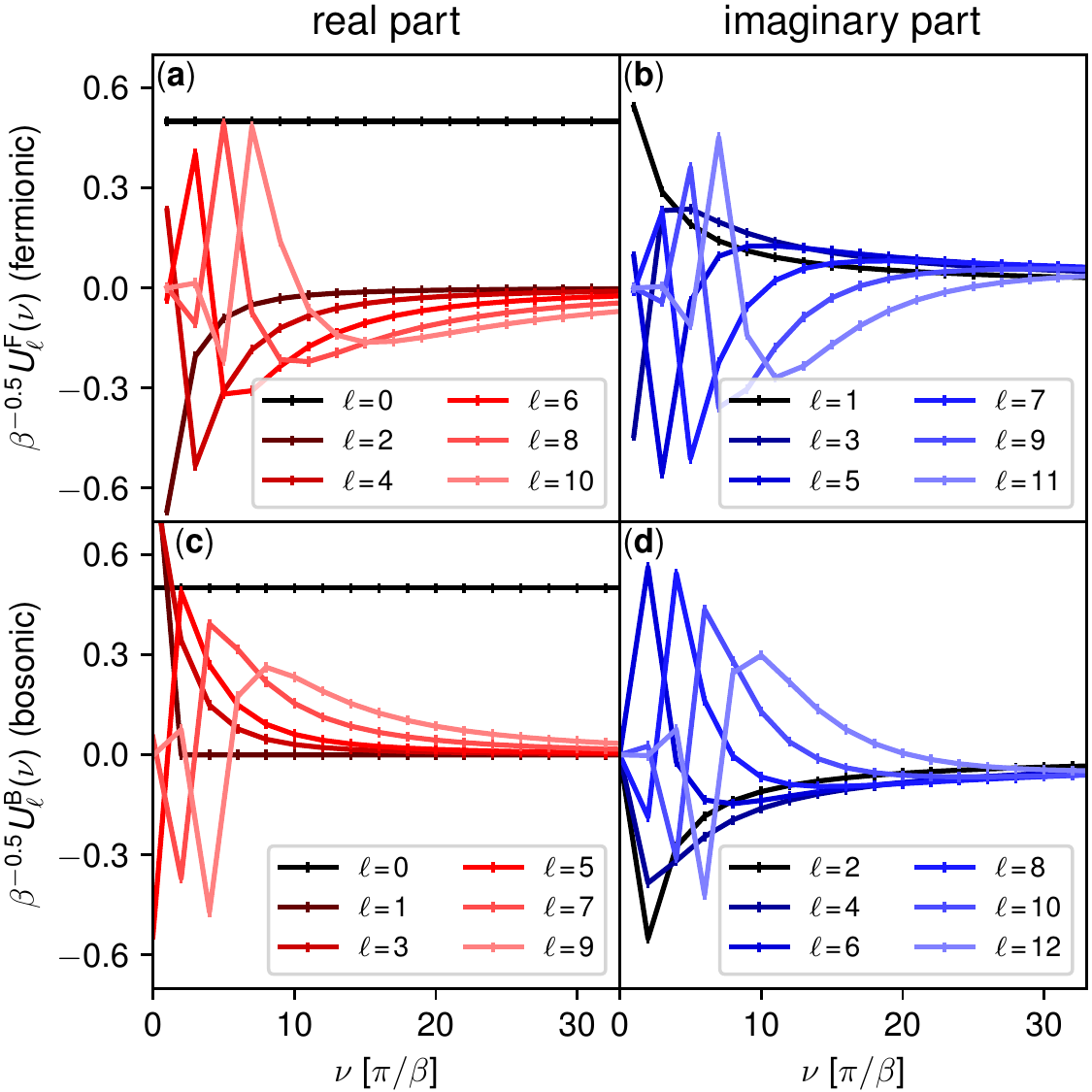}
    \caption{Lowest-order IR basis functions $\hat U_l(\iv)$:
    (a,b) for the augmented fermionic kernel (\ref{eq:KFbar}); (c,d) for the doubly augmented bosonic
    kernel (\ref{eq:KBtilde}).  Symmetry forces these functions to be either purely real (a,c) or
    purely imaginary (b,d) and we always plot the respective non-zero part.  The constant term $\hat U_0$
    is scaled by a factor 1/2.
    }
    \label{fig:Ulv}
\end{figure}

Neither vertex, $\hat M$ nor $\hat\Sigma$, can be modelled by the same basis as the Green's function, because they contain the Hartree--Fock term, which is a constant in frequency.  To model it we need to augment
the fermionic kernel in Eq.~(\ref{eq:friedholm}) to:
\begin{equation}
    K^\Fbar(\iv, \nu') = \frac1{\iv - \nu'} + S^\Fbar_0,
    \label{eq:KFbar}
\end{equation}
where $S^\Fbar_0$ is an arbitrary non-zero constant.  As the remaining terms in $\hat \Sigma$ and $\hat M$ behave like a scaled Green's function, one has truncated expansions analogous to Eq.~(\ref{eq:gtaurep}):
\begin{align}
    \hat \Sigma(\iv)&=\textstyle\sum_{l=0}^{L-1} \hat U^\Fbar_l(\iv) \Sigma_l + \epsilon_L,\label{eq:siwexp} \\
    \hat M(\iv)&=\textstyle\sum_{l=0}^{L-1} \hat U^\Fbar_l(\iv) M_l + \epsilon'_L,\label{eq:miwexp}
\end{align}
where $\Sigma_l$ and $M_l$ are again basis coefficients
and $\epsilon_L$ is an error term associated with truncating the series,
which is guaranteed to drop quickly.

In principle, $S^\Fbar_l$ and $\hat U^\Fbar_l$ are the singular values and left singular
functions, respectively, of the augmented kernel (\ref{eq:KFbar}). However, the kernel
is not compact, which means one is unable to compute the singular value expansion
numerically.  Instead, one chooses $\hat U^\Fbar_0(\iw) = 1$ and shifts the remaining terms
of the unaugmented kernel by one, i.e., $S^\Fbar_{l+1} = S^\FF_l$ and
$\hat U^\Fbar_{l+1} = \hat U^\FF_l$.  We plot the lowest order basis functions in Fig.~\ref{fig:Ulv}(a,b).

Fitting $\Sigma_l$ from Matsubara data for the self-energy requires us to choose a set of fitting frequencies.
Since the associated basis functions for $l=1,\ldots,L-1$ are identical to the underlying left singular functions,
the sampling frequencies $\WW^\FF$ for order $L' \ge L-1$ allow stable and compact fitting.  The
Hartree--Fock term $M_0$, on the other hand, is given by the limit $\iv\to\infty$, and corrections to this
asymptotic constant only decay as $1/\iv$.  Fitting $\Sigma_0$ from $\hat\Sigma(\iv)$ is thus a somewhat
delicate procedure:  on the one hand
we would like to have an extra sampling frequency for $\iv$ large, yet on the other hand one often has unfavorable
scaling of the uncertainty in $\hat\Sigma(\iv)$ with increasing frequency~\cite{Kaufmann_improved_estimators}, which discourages us from doing so.
We empirically observe that the distribution of $\WW^\FF$ for order $L'$ extends to higher frequencies as $L'$ is increased.
Therefore, a reasonable choice is to use $\WW^\FF$ for order $L'=L$ as $\WW^\Fbar$ for order $L$.

\begin{figure}
    \includegraphics[width=\columnwidth]{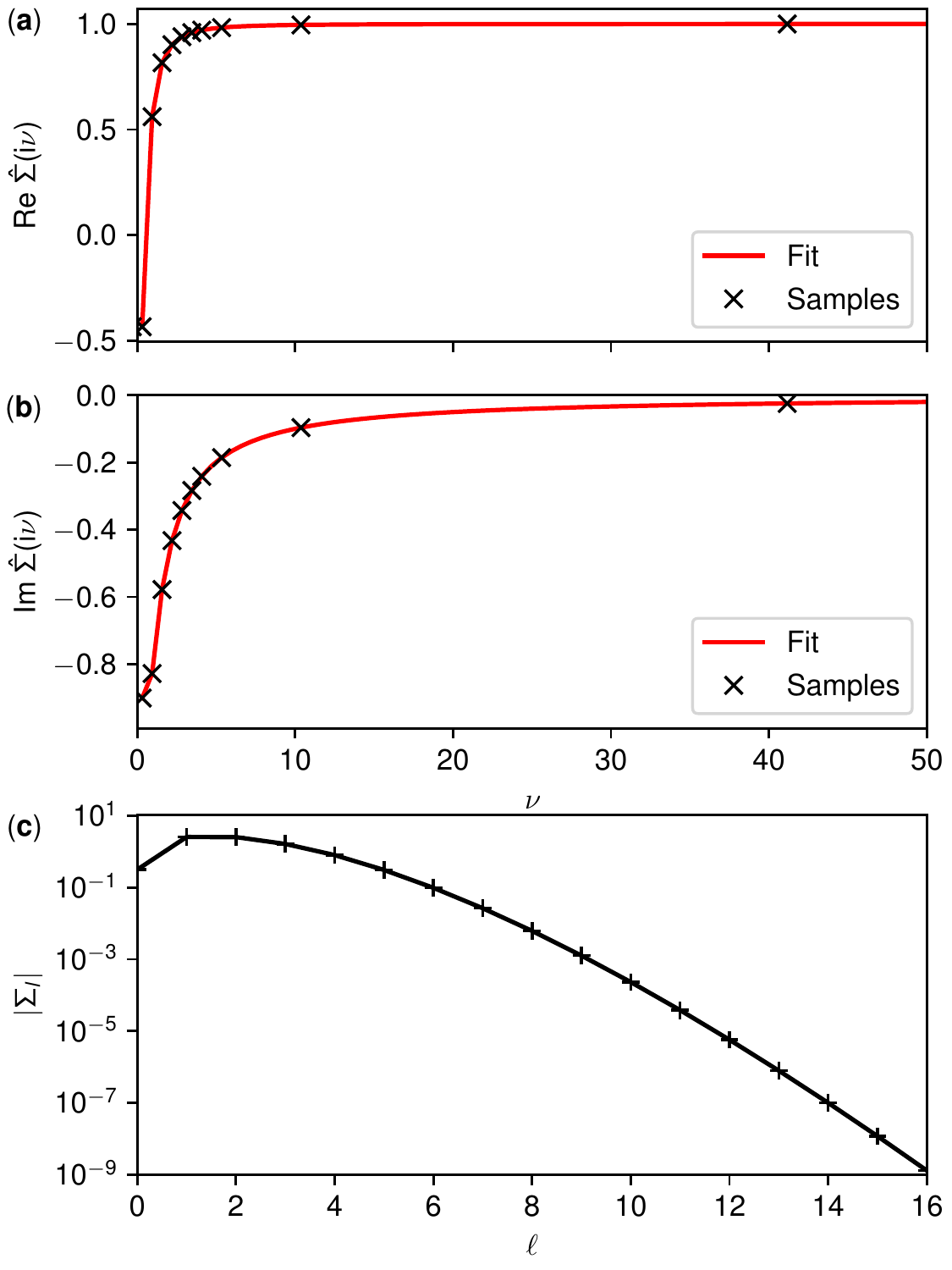}
    \caption{
    Expansion of the self-energy in the augmented fermionic basis: (a) Real part and (b) imaginary part of the self-energy (\ref{eq:self-energy-model}), (c) expansion coefficients (\ref{eq:siwexp}). 
    The crosses in (a) and (b) denote the data on the sampling frequencies.
    }
    \label{fig:self-energy}
\end{figure}

Let us demonstrate the fitting for a model self-energy:
\begin{align}
    \hat \Sigma(\iv) &= \Sigma_0 + \frac{1}{\iv-\epsilon_0},\label{eq:self-energy-model}
\end{align}
where $\Sigma_0 = 1$ and $\epsilon_0=0.5$ at $\beta=10$ ($\wmax=1$ and $S^\Fbar_L/S^\Fbar_0 \simeq 10^{-15}$).
The result is shown in Fig.~\ref{fig:self-energy}.
The augmented basis fits the self-energy accurately including the Hartree-Fock term $\Sigma_0$ from low to high frequencies.
As seen in Fig.~\ref{fig:self-energy}(c),
the expansion coefficient $\Sigma_l$ decay faster than exponentially~\footnote{At lower $T$, one may have to impose compactness explicitly on the expansion coefficients in the fitting as the augmented basis is overcomplete. See discussions in Sec. V.}.

We are now in a position to tackle a diagrammatic equation for the full one-particle vertex $\hat M$.
We note that by introducing the prefactor $(1 - \hat\Sigma\hat G_0)$, Eq.~(\ref{eq:dyson}) becomes a
linear equation for $\hat M$.  We can then insert Eq.~(\ref{eq:miwexp}) to arrive
at a least squares problem for the Dyson equation:
\begin{equation}
    \min_{M_l} \sum_{\iv\in\WW^\Fbar} \bigg| 
       \hat\Sigma(\iv) - \sum_{l=0}^{L-1} 
       \underbrace{\big[1 - \hat\Sigma(\iv)\hat G_0(\iv)\big] \hat U_l^\Fbar(\iv)}_{A_l(\iv)}
       M_l
     \bigg|^2.
\label{eq:sigmafit}
\end{equation}
Let us note that solving the Dyson equation by solving the least squares problem (\ref{eq:sigmafit})
is obviously not optimal: since Eq.~(\ref{eq:dyson}) is diagonal in frequency, one can
first solve the equation on the sampling frequencies and then fit $M_l$ from $M(\iv)$
in a second step~\cite{Li:2020eu}. However, in the two-particle case, this ceases to be an option,
since it involves convolutions over frequencies.

% -----------------------------------
\subsection{Bethe--Salpeter equation}
\label{sec:bse}
% -----------------------------------

After having developed the necessary tools for the sparse modeling of one-particle vertices and the rewriting of diagrammatic
equations as fitting problems for that case, we are ready to tackle the two-particle case.

\begin{figure}
    \centering
    \includegraphics[width=\columnwidth]{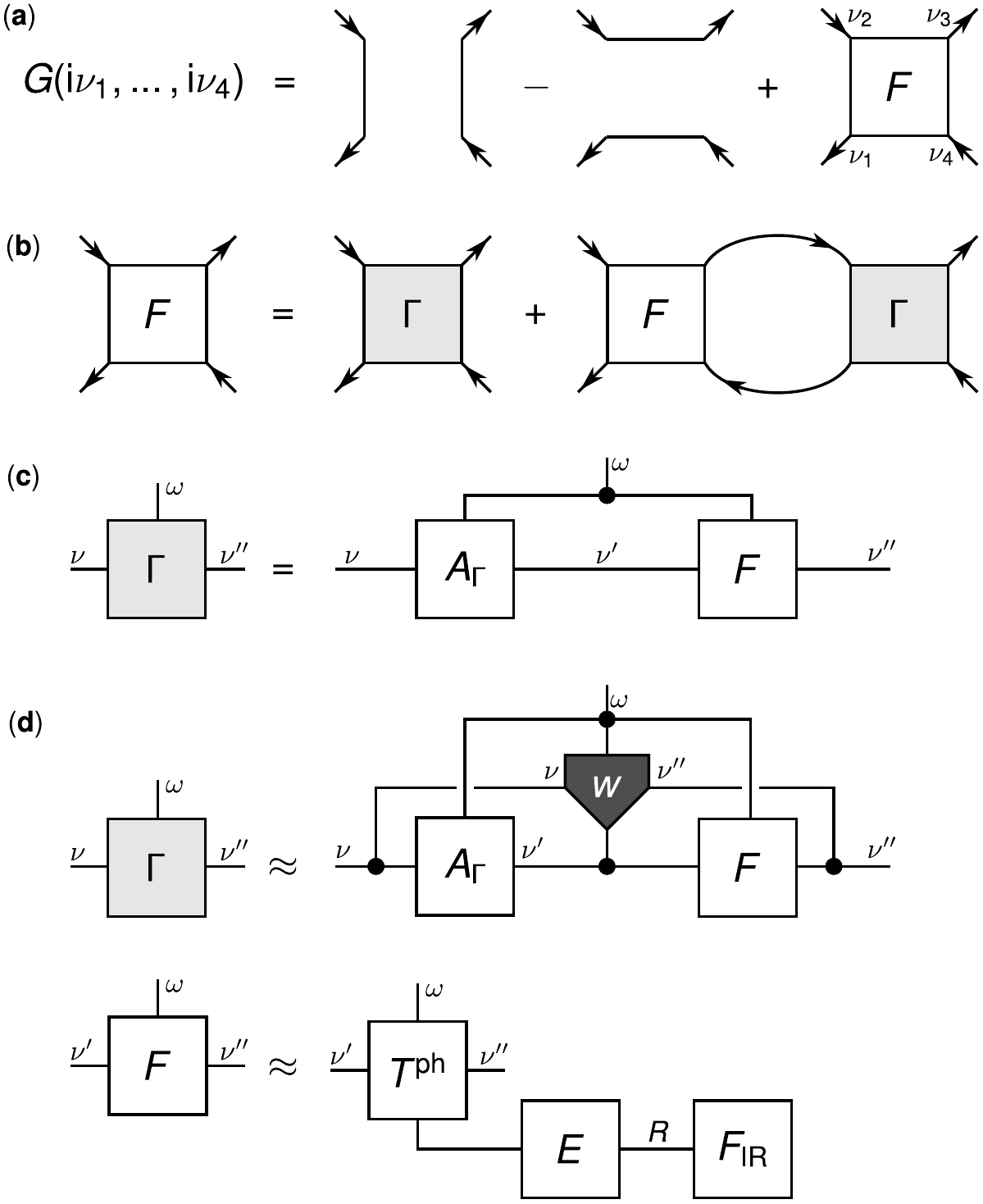}
    \caption{Sparse modeling for the Bethe--Salpeter equation:
    (a) diagrammatic representation the two-particle Green's function (\ref{eq:F}), where $F$ is the full two-particle vertex;
    (b) BSE in the particle--hole channel (\ref{eq:bse4}),
    where $\Gamma$ is the particle-hole irreducible two-particle vertex;
    (c) tensor network for the BSE in natural frequencies (\ref{eq:bse3}), where $A_\Gamma$ is defined in Eq.~(\ref{eq:AF});
    (d) sparse modeling and quadrature for the BSE (\ref{eq:bseir}), where $w$ is the tensor of summation weights,
    $T^\mathrm{ph}$ changes from the fermionic to the particle--hole frequency convention, $E$ is the expansion tensor
    (cf.~Fig.~\ref{fig:E}), and $F_\mathrm{IR}$ are the IR basis coefficients of the full vertex. \linethrough{$\bullet$} only gives a contribution if all connected indices have the same value.
    }
    \label{fig:bse}
\end{figure}

The two-particle analogue of Eq.~(\ref{eq:M}) reads:
\begin{equation}
\begin{split}
    & \hat G(\iv_1,\iv_2,\iv_3,\iv_4)= \beta^2\hat G(\iv_1) \hat G(\iv_3) \delta(\nu_1,\nu_2) \delta(\nu_3,\nu_4) \\
    & \quad-\ \beta^2\hat G(\iv_1) \hat G(\iv_3) \delta(\nu_1,\nu_4) \delta(\nu_3,\nu_2) \\
    & \quad+\ \hat G(\iv_1) \hat G(\iv_2) \hat F(\iv_1,\iv_2,\iv_3,\iv_4) \hat G(\iv_3) \hat G(\iv_4),
\end{split}
\label{eq:F}
\end{equation}
where $\hat F$ is the full two-particle vertex~\footnote{
Note that this convention is somewhat uncommon in literature, and especially differs from
Ref.~\onlinecite{RohringerRMP}.  There, usually a three-frequency convention is used.
}.
Diagrammatically, Eq.~(\ref{eq:F}) is shown in Fig.~\ref{fig:bse}(a).   There are now three different notions of two-particle reducibility in $F$: with respect to severing frequencies 1,2 from 3,4 (particle--hole channel), frequencies 1,4 from 3,2 (particle--hole transverse channel), and frequencies 1,3 from 2,4 (particle--particle channel). Consequently, there is an irreducible
vertex and a corresponding diagrammatic equation for each of these channels.\cite{Bickers04,RohringerRMP}

Without loss of generality, we will restrict ourselves to the particle--hole channel for now.  The Bethe--Salpeter equation, which relates the full vertex $\hat F$ to the irreducible vertex $\hat\Gamma$, reads [cf. Fig.~\ref{fig:bse}(b)]:
\begin{equation}
\begin{split}
    &\hat F(\iv_1,\iv_2,\iv_3,\iv_4) = \hat\Gamma(\iv_1,\iv_2,\iv_3,\iv_4) \\
    &\!+ \frac1{\beta^2}\sum_{\nu,\nu'}\hat F(\iv_1,\iv_2,\iv,\iv')
    \hat G(\iv) \hat G(\iv') \hat\Gamma(\iv',\iv,\iv_3,\iv_4).
\end{split}
\label{eq:bse4a}
\end{equation}
Solving the BSE in this convention is cumbersome, as it requires a sum over two
inner frequencies, one of which is fixed by conservation of energy.  This can be eliminated
by switching into the ``natural'' frequency convention for the particle--hole channel:
\begin{equation}
    \hat F(\iw; \iv, \iv') := \frac{1}{\beta} \hat F(\iv+\iw,\iv',\iv'-\iw,\iv),
    \label{eq:ph}
\end{equation}
where $\iw$ is a bosonic transfer frequency and $\iv,\iv'$ are fermionic
frequencies, and from now on the three-argument version of any quantity indicates
the particle--hole convention~\footnote{Note, that there is an additional prefactor $\beta$ coming from  the evaluation of a $\delta$ for enforcing the energy conservation.}.  Equation~(\ref{eq:bse4a}) then reads:
\begin{equation}
\begin{split}
& \hat F(\iw;\iv',\iv'')= \hat\Gamma (\iw;\iv,\iv'') \\ 
&\quad+ \frac 1\beta \sum_{\nu'} \hat \Gamma(\iw;\iv,\iv')\,\hat G(\iv')\,\hat G(\iv'+\iw)\, \hat F(\iw;\iv',\iv'').
  \label{eq:bse4}
  \end{split}
\end{equation}
This equation can be rewritten into the following form:
\begin{equation}
  \hat\Gamma (\iw;\iv,\iv'') = \frac 1\beta \sum_{\nu'} A_\Gamma(\iw;\iv,\iv')\,\hat F(\iw;\iv',\iv''),
  \label{eq:bse3}
\end{equation}
where we have defined a ``Bethe--Salpeter operator'' $A_\Gamma$:
\begin{equation}
    A_\Gamma(\iw;\iv,\iv')= \beta\delta_{\nu\nu'} - \hat \Gamma(\iw;\iv,\iv')\,\hat G(\iv')\,\hat G(\iv'+\iw).
\label{eq:AF}
\end{equation}

\begin{figure*}
    \centering
    \includegraphics[width=\textwidth]{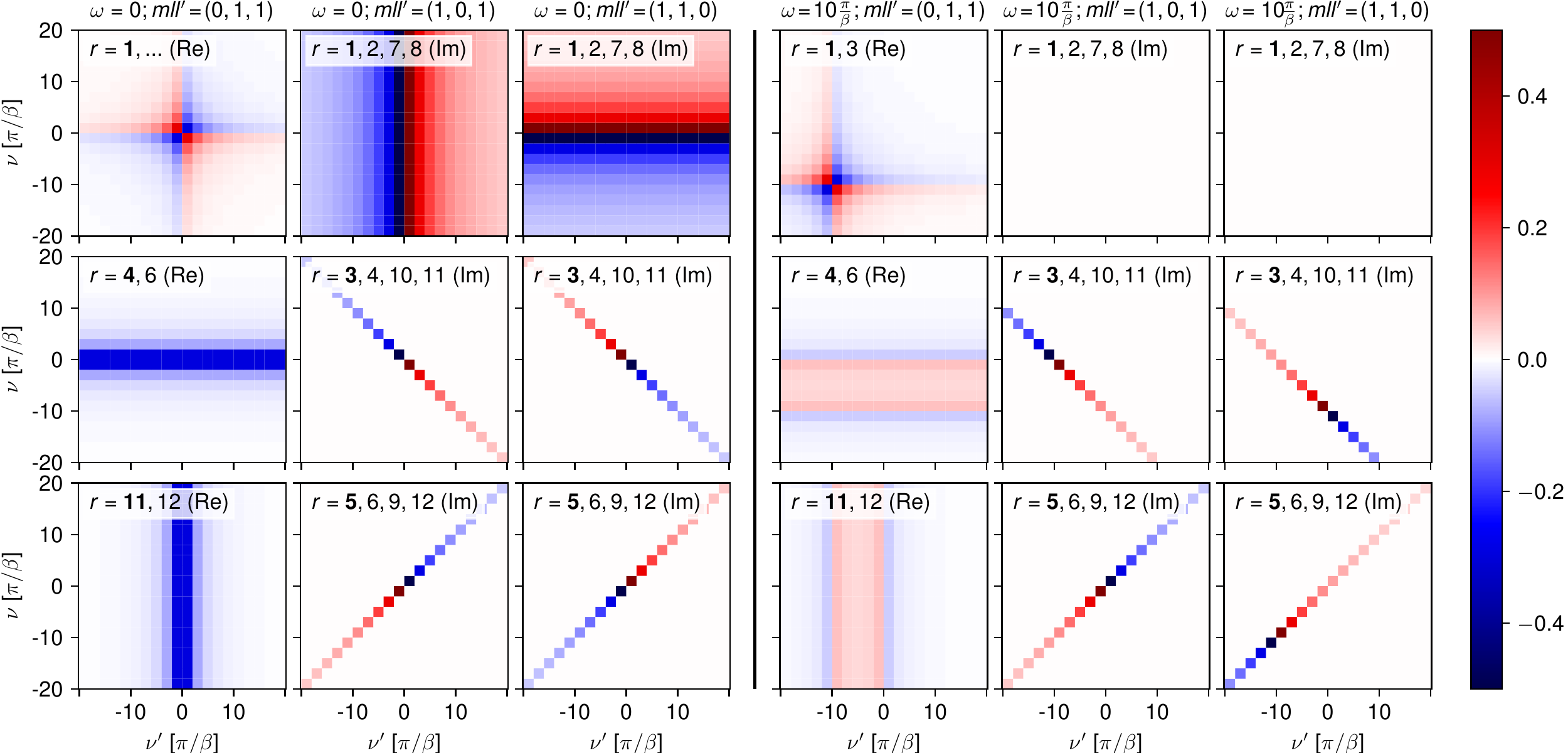}
    \caption{Four-point vertex basis functions $\hat U_l^\Fbar \hat U_{l'}^\Fbar \hat U_m^\Btilde$ for selected $r,l,l',m$, in Matsubara frequencies and projected back into the particle--hole convention $(i\omega,i\nu,i\nu')$. We plot over a fermionic frequency box $(\nu,\nu')$
    for two given bosonic frequencies: $\omega=0$ (left side) and
    $\omega = 10\pi/\beta$ (right side).  Columns show different
    expansion orders $(l,l',m)$, while the rows show different representations
    $r$ (the first number, indicated in bold, is the representation plotted; other numbers given denote structurally similar terms). Symmetry dictates that the
    functions must be purely real (Re) or imaginary (Im), and we plot only the
    respective part.
    }
    \label{fig:low_order_vertex}
\end{figure*}

\subsubsection{modeling of the two-particle vertex.}
Analogous to the expansion of the
two-particle Green's function~(\ref{eq:g4ir}), we expand the two-particle vertex in an overcomplete basis:
\begin{equation}
\begin{split} 
&\hat{F}(\iv_1,\ldots,\iv_4)=\sum_{r=1}^{12} \sum_{\nu\nu'\omega} T_r(\iv_1,\ldots,\iv_4; \iv, \iv', \iw) \\
&\qquad\times\sum_{l,l',m=0}^{L-1} \hat U_l^\Fbar(\iv) \hat U_{l'}^\Fbar(\iv') \hat U_m^\Btilde(\iw) F_{r,ll'm} + \epsilon_L,\\
\end{split}
\label{eq:f4ir}
\end{equation}
and similarly for the irreducible vertex $\Gamma$.

Like the self-energy, $F$ and $\Gamma$ have an asymptotically constant background.
Therefore, as in Eq.~(\ref{eq:miwexp}), the fer\-mi\-o\-nic kernel has to be augmented using Eq.~(\ref{eq:KFbar}).   At the same time, the bosonic kernel for the two-particle Green's function $K^\Bbar$ has to be further augmented using a bosonic analogue of Eq.~\eqref{eq:KFbar}.
This defines an augmented bosonic kernel $K^\Btilde$:
\begin{equation}
\begin{split} 
    K^\Btilde&(\iw, \omega') = K^\Bbar(\iw,\omega') + S_0^\Btilde  \\
    &= \frac{\omega'}{\iw-\omega'}  + S_0^\Btilde + S^\Btilde_{1} \delta_{\iw,0} + \frac{S^\Btilde_2 (1-\delta_{\iw,0})}{\iw}. \\
\end{split}
\label{eq:KBtilde}
\end{equation}
(We again shift the remaining singular values by $S^\BB_l \to S^\Btilde_{l+3}$).  We plot the
corresponding one-particle basis functions in Fig.~\ref{fig:Ulv}(c,d).

In addition to the background, the vertex has a rich asymptotic structure, consisting of
``lines'' and ``planes'' running horizontally, vertically and diagonally through the frequency
domain and extending to infinity~\cite{Rohringer2012}.  These structures are captured by
the augmented basis functions when combined with the frequency translations in the different representations $r$.
For $F$, we can prove this by applying the same arguments as for
$\Sigma$ (cf.~Sec.~\ref{sec:dyson}), but for the dependence on each of the outer frequencies $\iv_1,\ldots,\iv_4$ separately. $F$ differs from the connected two-particle propagator  (\ref{eq:F})
by removal of four one-particle Green's function lines (one for each outer frequency).  Similar to $\Sigma$,  the dependence of $F$ on each of the frequencies then amounts to
a constant term plus a scaled one-particle Green's function, translated through the frequency translation tensor $T_r$. 

For the irreducible vertex $\Gamma$, one can show that the
asymptotic part has a similar fundamental form as for $F$ away from the divergence lines
(cf.~Sec.~\ref{sec:atom}). We thus conjecture that the expansion (\ref{eq:F}) remains compact also for $\Gamma$, and offer numerical evidence in Secs.~\ref{sec:bench} and \ref{sec:impurity}.

Let us illustrate the effect of the augmented two-particle basis by plotting the low-order basis functions
in Fig.~\ref{fig:low_order_vertex} in Matsubara frequencies in the particle--hole convention, cf. Eq.~(\ref{eq:ph}).  The results for $\omega=0$ as given on the left side.  We see that, e.g., $F_{r,110}$ translates to the horizontal line structure for $r = 4$ and to the vertical line for $r = 11$.  As expected,
these structures shift away from the center for $\omega = 10$ (right side).

\subsubsection{Solving the Bethe-Salpeter equation}

Eq.~(\ref{eq:bse3}) is a set of linear equations which can be solved
independently for each $\iw$.  However, unlike Eq.~(\ref{eq:dyson}),
it involves the (infinite) sum over an inner fermionic line [cf.~Fig.~\ref{fig:bse}(c)].
As vertices, $F$ and $\Gamma$ are asymptotically constant,
which means the summand in Eq.~(\ref{eq:bse3}) decays only by virtue
of the Green's function lines in Eq.~(\ref{eq:AF}) as $O(1/\nu'^2)$.
Asymptotic expansions of the vertex~\cite{Wentzell20, Krien19,Li16,Kunes11} improve upon this,
but require additional knowledge of a set of two- and three-point correlation functions, together with
a (usually uncontrolled) method of connecting asymptotic and non-asymptotic region.

Storing enough frequencies to reliably perform the fermionic sum
quickly exceeds available memory.
Fortunately, using sparse modeling
(Sec.~\ref{sec:sparse4}) we can solve this: analogous to
Eq.~(\ref{eq:sigmafit}),
we rewrite Eq.~(\ref{eq:bse3}) as a
least squares problem for the IR coefficients $F_{r,ll'm}$:
\begin{equation}
\begin{split}
    \min_F&\sum_{(\iw,\iv,\iv'')\in\WW'}\Big|\Gamma(\iw;\iv,\iv'') - \frac 1\beta \sum_{\nu'}\sum_{r,ll'm} \\
    &\times A_\Gamma(\iw;\iv,\iv')\,E_{r,ll'm}^{\iv'+\iw,\iv'',\iv''-\iw,\iv'} F_{r,ll'm}\Big|^2.
\end{split}
\label{eq:bse4ir-inf}
\end{equation}
Due to the augmentated kernels entering Eq.~(\ref{eq:f4ir}), $\WW'$ is the set
of sampling frequencies generated from Eq.~(\ref{eq:wsample4}), but
with $\WW^\FF$ ($\WW^\Bbar$) replaced by $\WW^\Fbar$ ($\WW^\Btilde$) and transformed in the particle-hole basis.

Eq.~(\ref{eq:bse4ir-inf}) solves the storage problems, but it still
requires a sum over an infinite set of frequencies.  A naive truncation
of that sum to the innermost $N$ frequencies will only converge as
$\bigO(1/N)$.  In Sec.~\ref{sec:convolution}, we will improve on this
by developing an algorithm which replaces the full infinite sum by a weighted
finite sum:
\begin{equation}
\begin{split}
    &\frac1\beta \sum_{\nu'} A_\Gamma(\iw;\iv,\iv')\,\hat F(\iw;\iv',\iv'') \\
    &\quad\approx\!\!\!\sum_{\nu'\in\mathcal \WW_{\omega\nu\nu''}}\!\!\!\!w_{\omega\nu\nu''}(\iv')\,A_\Gamma(\iw;\iv,\iv')\,\hat F(\iw;\iv',\iv''),
\end{split}
\label{eq:conv}
\end{equation}
where $\mathcal \WW_{\omega\nu\nu''}$ are the set of quadrature frequencies for the outer 
frequencies $\iw,\iv,\iv''$ and $w_{\omega\nu\nu''}(\iv')$ are the corresponding integration weights. 
The summation rule in 
Eq.~(\ref{eq:conv}) is observed to converge exponentially to the true result with respect to the
number of quadrature points.

Inserting Eq.~(\ref{eq:conv}) into Eq.~(\ref{eq:bse4ir-inf}), we arrive at:
\begin{equation}
    \begin{split}
    \min_F& \sum_{(\iw,\iv,\iv'')\in\WW'}\Big|
    \Gamma(\iw;\iv,\iv'') - \sum_{\nu'\in\mathcal \WW_{\omega\nu\nu''}}\!\!w_{\omega\nu\nu''}(\iv') \\
    &\times \sum_{r,ll'm} A_\Gamma(\iw;\iv,\iv')\,E_{r,ll'm}^{\iv'+\iw,\iv'',\iv''-\iw,\iv'} F_{r,ll'm}\Big|^2,
\end{split}
\label{eq:bseir}
\end{equation}
where we have omitted a, usually necessary, regularization term for brevity, cf. Eq.~(\ref{eq:g4fit}).
Diagrammatically, Eq.~(\ref{eq:bseir}) is shown in Fig.~\ref{fig:bse}(d). By using this equation,
we have both the truncation error and the quadrature error under control.  Therefore we can expect
exponential convergence in $F$ in both time and memory.  We will discuss details of the
fitting in Sec.~\ref{sec:expansion}.

Let us note that the other direction of the Bethe--Salpeter equation (\ref{eq:bse4}), 
i.e., computing $\Gamma$ from $F$, can be done by introducing:
\begin{equation}
    A_F(\iw;\iv,\iv')= \beta\delta_{\nu\nu'} + 
    \hat F(\iw;\iv,\iv')\,\hat G(\iv')\,\hat G(\iv'+\iw),
    \label{eq:Aother}
\end{equation}
and switching $F$ and $\Gamma$ in Eqs.~(\ref{eq:bse3}) to (\ref{eq:bseir}).

% =============================================
\section{Sparse convolution}
\label{sec:convolution}
% =============================================

Ultimately, we want to perform convolutions of vertices in a specific
channel, cf.~Eq.~(\ref{eq:bse3}), which requires an (infinite) sum over a fermionic
Matsubara frequency.  As outlined in Sec.~\ref{sec:bse}, truncating this sum to the
innermost $N$ frequencies will only converge as $\bigO(1/N)$, which is why we are seeking
to replace it with a weighted sum over a finite set of frequencies instead, cf.~Eq.~(\ref{eq:conv}).

% ---------------------------
\subsection{Simpler case: the Lindhard bubble}
\label{sec:lindhard}
% ---------------------------

To motivate our sparse convolution scheme, let us first consider a simpler problem, which we will
again use to develop the necessary tools: Let $A$ and $B$ be one-particle Green's functions and
let $C$ be defined as
\begin{equation}
C(\iw)=\frac1\beta\sum_{\nu}A(\iv)B(\iv+\iw),
\label{eq:prod}
\end{equation}
where $\iw$ is a bosonic or fermionic Matsubara frequency and, correspondingly
$B$ and $C$ are bosonic or fermionic Green's functions.  The sum converges without a convergence factor as $A(\iv)B(\iv+\iw)$ decays as $\bigO(1/\nu^2)$. Using the residual
calculus, it is straightforward to show that the product (\ref{eq:prod})
can be decomposed as follows:
\begin{equation}
  A(\iv)B(\iv+\iw) = A'_\omega(\iv) + B'_\omega(\iv+\iw),
  \label{eq:absplit}
\end{equation}
where $A'$ and $B'$ are families of auxiliary Green's functions in
$\iv$ and $\iv+\iw$, respectively, and both families are parameterized by
$\iw$.  (For completeness, we show this relation in Appendix~\ref{app:residual}.)

Equation~(\ref{eq:absplit}) again admits an overcomplete representation of the integrand
in terms of two sets of coefficients, $A'_{\omega,l}$ and $B'_{\omega,l}$:
\begin{equation}
    A(\iv)B(\iv+\iw) = \sum_{l=0}^{L-1} \big[ \hat U^\FF_l\!(\iv) A'_{\omega,l} + \hat U^\alpha_l(\iv+\iw) B'_{\omega,l} \big] + \epsilon_L.
    \label{eq:ABexpand}
\end{equation}
Since each constituent can be represented compactly with the IR basis, there
exists a compact representation for the product and the error drops
superexponentially, $\epsilon_L \sim S_l$. Using Eq.~(\ref{eq:ABexpand}), we
can compute the Matsubara sum:
\begin{equation}
    C(\iw) = \sum_{l=0}^{L-1} [ U^\FF_l(0^-) A'_{\omega,l} + U^\alpha_l(0^-) B'_{\omega,l} ] + \epsilon_L,
    \label{eq:CsumU}
\end{equation}
where $U^\alpha_l(0^-)$ are the Fourier transform of the $l$-th bosonic or fermionic
basis function $\hat U_l$, evaluated at $\tau = 0^-$.

We proceed in a manner similar to the overcomplete representation of the
two-particle function, but for each value of $\iw$ separately.
We first generate a set $\WW_\omega$ of fermionic sampling frequencies for
$A(\iv)B(\iv+\iw)$: expanding $A'$ in $\iv$ corresponds to the standard set $\WW_A=\WW^\FF$ of
fermionic sampling frequencies (\ref{eq:wsample2}), and expanding $B'$ in $\iv+\iw$ corresponds to a shifted set
$\WW_B$ of fermionic sampling frequencies.
The full set is then just the
union of both individual sets:
\begin{equation}
    \WW_\omega = \WW_A \cup \WW_B = \WW^\FF \cup \{\iv-\iw:\iv\in \WW^\alpha \}.
    \label{eq:VAB}
\end{equation}
Using the frequency set (\ref{eq:VAB}), we can turn Eq.~(\ref{eq:ABexpand}) into
a least squares problem:
\begin{equation}
    \min_{A',B'} \left\Vert [AB]_\omega 
    - \begin{pmatrix} \hat U^\FF_0 & \hat U^\alpha_\omega \end{pmatrix} \begin{pmatrix} A'_\omega \\ B'_\omega \end{pmatrix}
    \right\Vert,
    \label{eq:fitABmodel}
\end{equation}
where the data vector is $[AB]_{\omega,i}=A(\iv_i)B(\iv_i+\iw)$, $\iv_i$
runs over frequencies in $V_\omega$, the design matrix
is given block-wise as $[\hat U^\alpha_\omega]_{il} = \hat U^\alpha_l(\iv_i + \iw)$, and the fitting vector is just
the IR coefficients of $A'$  and $B'$, stacked vertically.
The Matsubara sum (\ref{eq:CsumU}) is then given by:
\begin{equation}
    C(\iw) = \sum_{\iv\in V_\omega} w_\omega(\iv) A(\iv) B(\iv + \iw),
\end{equation}
i.\,e., the full sum is replaced by a weighted sum over a small subset
of frequencies.  The vector of integration weights $w$ is determined by the
solution of the following least squares problem:
\begin{equation}
    \min_{w_\omega} \left\Vert
    \begin{pmatrix} u^\FF \\ u^\alpha \end{pmatrix}
    -
    \begin{pmatrix} [U^\FF_0]^\mathrm{T}  \\ [\hat U^\alpha_\omega]^\mathrm{T}  \end{pmatrix}
    w_\omega
    \right\Vert,\label{eq:fit-weight}
\end{equation}
where $w_{\omega,i}=w_\omega(\iv_i)$, the evaluation vector
is given block-wise as $[u^\alpha]_l=U^\alpha_l(0^-)$, and where $\hat U^\mathrm{T}$ denotes the
transpose of the design matrix in Eq.~(\ref{eq:fitABmodel}).
If Eq.~(\ref{eq:fit-weight}) is underdetermined, we take its least norm solution.

% ---------------------------
\subsection{Full two-particle convolution}
\label{sec:fourpoint}
% ---------------------------
Now we turn to the case (\ref{eq:bse3}) of multiplying two two-particle
functions.  For simplicity, we focus on the particle--hole channel.
Similar relations can be inferred for the transverse channel and for
the particle--particle channel.

By transforming Eq.~(\ref{eq:g4lehmann}) into the particle--hole
convention through Eq.~(\ref{eq:ph}) and focussing on the dependence on
$\iv'$, one has:
\begin{equation}
\begin{split}
A(\iw;\iv,\iv') &
= A^{(1)}_{\omega\nu}(\iv') + A^{(2)}_{\omega\nu}(\iv'+\iw) \\
&+ A^{(3)}_{\omega\nu}(\iv'-\iv) + A^{(4)}_{\omega\nu}(\iv'+\iv+\iw),
\end{split}
\label{eq:AFquad}
\end{equation}
where $A^{(1)}$ to $A^{(4)}$ are a family of auxiliary objects. With 
$\iw$ and $\iv$ held fixed, $A^{(1)}$ and $A^{(2)}$ have the structure
of a fermionic Green's function, while $A^{(3)}$ and $A^{(4)}$ are bosonic
Green's functions.  Similarly, for the dependence on the other fermionic
frequency, one finds:
\begin{equation}
\begin{split}
F(\iw;\iv',\iv'') &
= F^{(1)}_{\omega\nu''}(\iv') + F^{(2)}_{\omega\nu''}(\iv'+\iw) \\
&+ F^{(3)}_{\omega\nu''}(\iv'-\iv'') + F^{(4)}_{\omega\nu''}(\iv'+\iv''+\iw).
\end{split}
\label{eq:Gammaquad}
\end{equation}

Similar to Eq.~(\ref{eq:absplit}), we insert  Eqs.~(\ref{eq:AFquad}) and (\ref{eq:Gammaquad}) into
the Eq.~(\ref{eq:bse3}) and use the residual calculus to obtain a model for the summand:
\begin{equation}
\begin{split}
&A(\iw;\iv,\iv') F(\iw;\iv',\iv'') \\
&\quad= X^{(1)}_{\omega\nu\nu''}(\iv') + X^{(2)}_{\omega\nu\nu''}(\iv'+\iw) \\
&\quad+ X^{(3)}_{\omega\nu\nu''}(\iv'-\iv) + X^{(4)}_{\omega\nu\nu''}(\iv'+\iv+\iw)\\
&\quad+ X^{(5)}_{\omega\nu\nu''}(\iv'-\iv'') + X^{(6)}_{\omega\nu\nu''}(\iv'+\iv''+\iw),
\end{split}
\label{eq:prodmodel}
\end{equation}
where $X^{(1)}$, \ldots, $X^{(6)}$, with $\omega,\nu,\nu''$ held fixed, are again 
Green's functions.

This means we can generate an overcomplete representation consisting of six terms, and
the set of sampling frequencies becomes:
\begin{equation}
\begin{split}
    \WW_{\omega\nu\nu''} &= \big\{\iv'+\iw_\mathrm{s}: \iv'\in\WW^\mathrm{F}, \iw_\mathrm{s}\in\{0, -\iw\} \big\} \\
    &\,\cup\big\{\iv'+\iv_\mathrm{s}: \iv'\in\WW^\Bbar,  \\
    &\hspace{5.9em}\iv_\mathrm{s}\in\{\iv, \iv'', -\iv-\iw, -\iv''-\iw\} \big\},
\end{split}
\label{eq:wquad}
\end{equation}
where the ``mixing'' of fermionic and bosonic models in Eq.~(\ref{eq:prodmodel}) is
reflected in the presence of both fermionic and bosonic sampling frequencies, shifted
by a bosonic and fermionic shift frequency, $\iw_\mathrm{s}$ and $\iv_\mathrm{s}$,
respectively, to create a grid of fermionic frequencies~\footnote{
Let us emphasize that vertex augmentation of the basis, cf. Sec.~\ref{sec:dyson},
is not necessary here for the quadrature as the summand behaves as a Green's function
like by virtue of Eq.~(\ref{eq:bse3}).
}.

This fixes the quadrature frequencies in Eq.~(\ref{eq:conv}).  What remains to be
determined are the weights.  Analogous to Eq.~(\ref{eq:fit-weight}), $w_{\omega\nu\nu''}$
is given by the solution to the least squares problem:
\begin{equation}
    \min_{w_{\omega\nu\nu''}} \left\Vert
    \begin{pmatrix} u^\mathrm{F} \\ u^\mathrm{F} \\ u^\Bbar \\ \vdots \\ u^\Bbar \end{pmatrix}
    -
    \begin{pmatrix} [U^\mathrm{F}]^\mathrm{T} \\ [U^\mathrm{F}_\omega]^\mathrm{T} \\ 
                    [U^\Bbar_{\nu}]^\mathrm{T} \\ \vdots \\ [U^\Bbar_{\iv''+\iw}]^\mathrm{T}  \end{pmatrix}
    w_{\omega\nu\nu''}
    \right\Vert,\label{eq:wquad-fit}
\end{equation}
where again the evaluation vector is given block-wise by $[u^\alpha]_l=U^\alpha_l(0^-)$
and the design matrix is formed using blocks of $[\hat U^\alpha_\omega]_{il} = \hat U^\alpha_l(\iv_i + \iw)$.

\begin{figure}
    \centering
    \includegraphics[width=\columnwidth]{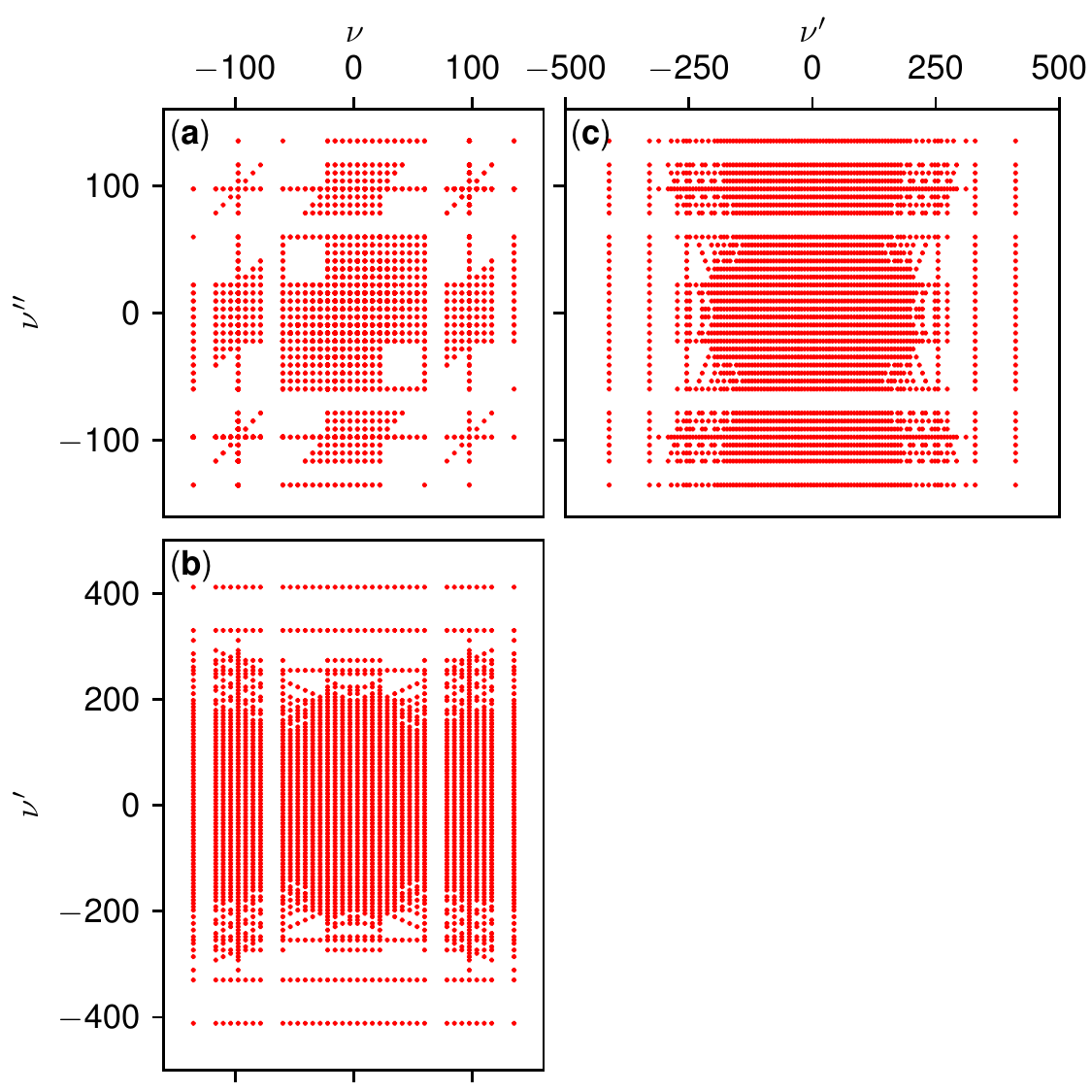}
    \caption{
Sparse frequency sets for $\beta=1$ and $\wmax=10$ with a cutoff of $10^{-5}$:
(a) sparse sampling frequency set $\WW'$, (b) left-side quadrature frequencies $\WW_\mathrm{L}$
for the BSE, (c) right-side quadrature frequencies $\WW_\mathrm{R}$.  We show only
points for zero bosonic frequency, $\iw = 0$.
    }
    \label{fig:quadrature}
\end{figure}

Figure~\ref{fig:quadrature}(a) shows sampling frequencies $\WW'$ for a vertex function for $\beta=1$ and $\wmax=10$ with a cutoff of $10^{-5}$.
We define quadrature points for the left and right objects as $\WW_\mathrm{L}\equiv \big\{(\iv, \iv', \iw): \iv'\in \WW_{\omega\nu\nu''}\land (\iv,\iv'',\iw) \in \WW'\big\}$
and  $\WW_\mathrm{R} \equiv \big\{(\iv', \iv'', \iw): \iv'\in \WW_{\omega\nu\nu''}\land (\iv,\iv'',\iw) \in \WW'\big\}$, respectively.
We plot $\WW_\mathrm{L}$ and $\WW_\mathrm{R}$ in  Figs.~\ref{fig:quadrature}(b) and \ref{fig:quadrature}(c), respectively.
The sampling frequencies and quadrature points are distributed sparsely especially at high frequencies.

With both quadrature points and weights specified, let us discuss computational cost and scaling.
From Eq.~(\ref{eq:wquad}), we have that for each choice of ``outer'' frequencies,
$2L \le |\WW_{\omega\nu\nu''}| \le 6L$, with values for typical outer frequencies close to
$|\WW_{\omega\nu\nu''}|\approx 6L$.  The size of the design matrix in Eq.~(\ref{eq:wquad-fit}) is
$6L \times |\WW_{\omega\nu\nu''}|$, thus the weights require $\bigO(L^3)$ time to compute for
each outer frequency. 

In solving the BSE (\ref{eq:bseir}), the quadrature has to be computed for each of the
sampling frequencies in $\WW$ (\ref{eq:wsample4}).  Since one has $|\WW|\approx 8L^3$
(cf.~Sec.~\ref{sec:sparse4}), this implies we in total have to store $\approx 50L^4$ quadrature
weights and frequencies.  The weights should be precomputed at a cost of $\bigO(L^6)$.
The quadrature (convolution) then takes $\bigO(L^4)$ time to compute, as it is merely a
weighted sum over $\bigO(L)$ frequencies for each of the $\bigO(L^3)$ sampling frequencies.

% =============================================
\section{Basis expansion and fitting}
\label{sec:expansion}
% =============================================
We will now discuss the solution of the least squares problem (\ref{eq:g4fit}). For this, it is useful
to first ``flatten'' the tensor $E$ into a matrix form.  We thus impose an ordering on the sampling
frequency set (\ref{eq:wsample4}) and on the set of basis coefficients:
\begin{align}
    \WW &= \big\{(\iv_{1V}, \iv_{2V}, \iv_{3V}, \iv_{4V})\big\}_{V=1,\ldots,|\WW|},\label{eq:Wset} \\
    \mathcal R &= \big\{(r_R, l_R, l'_R, m_R)\big\}_{R=1,\ldots,|\mathcal R|},\label{eq:Rset}
\end{align}
where $V$ is an index into the sampling point grid and $R$ is a flat index into $r,ll'm$.  Correspondingly,
we define $\hat G_V:=G(\iv_{1V},\ldots,\iv_{4V})$ and $G_R:=G_{r_R,l_R,l^\prime_R,m_R}$. With these
definitions, we arrive at the matrix form of Eq.~(\ref{eq:g4fit}):
\begin{equation}
    \min_\rho \Big\Vert \hat G_V - \sum_{R\in\mathcal R} E_{VR} G_R \Big\Vert^2 + \lambda \Big\Vert\Gamma_R G_R\Big\Vert^2,
    \label{eq:g4matrixfit}
\end{equation}
where $E_{VR}$ is the flattened version of $E^{\nu_1\nu_2\nu_3\nu_4}_{r,ll'm}$ (we will discuss its
explicit form shortly).

After constructing the matrix $E_{VR}$, Eq.~(\ref{eq:g4matrixfit}) can be fed directly to a ordinary least
squares solver.  However, for large $L$, the cost can be prohibitive: from Eq.~(\ref{eq:Rset}), one has
$|\mathcal R|=12 L^3$.  Constructing $E$ thus requires storing $\approx 100 L^6$ numbers and solving
the least squares problem requires $\bigO(L^9)$ flops.  Even though one has $L=\bigO(\log(\beta\wmax\epsilon^{-1}))$,
this will only be viable for small values of $L$.

\begin{figure}
    \centering
    \includegraphics[scale=0.7]{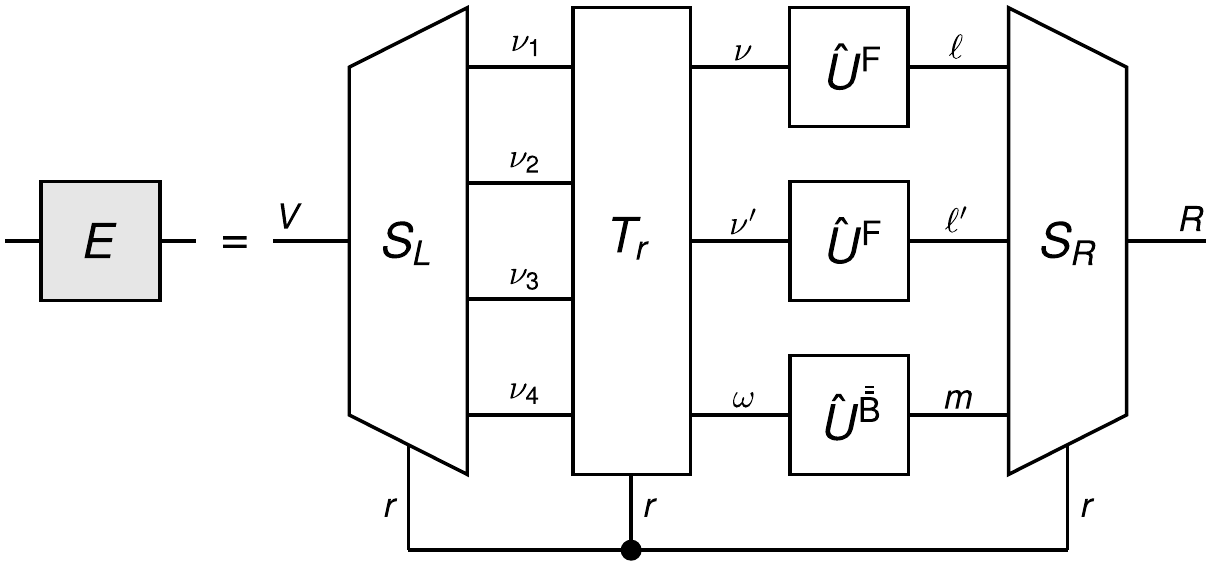}
    \caption{Tensor network representation of the expansion matrix $E$ (\ref{eq:E}). $S_L$ and $S_R$ are tensors which select
    indices from the inner side and ``flatten'' them to a single index, $T_r$ is the frequency translation tensor
    (cf.~Table~\ref{tab:transl}), $\hat U^\alpha$ are one-particle transformation matrices from IR to frequencies
    with their elements given by Eq.~(\ref{eq:sve}), and \linethrough{$\bullet$} only gives a contribution if all
    connected indices have the same value.
    }
    \label{fig:E}
\end{figure}

For larger $L$, we would like to construct $E G$ and $E^\dagger \hat G$ on the fly and use an
iterative least squares solver.  We start with the explicit form of $E$
by combining Eqs.~(\ref{eq:g4ir}) and (\ref{eq:g4matrixfit}):
\begin{equation}
\begin{split}
    E_{VR} = \sum_{\nu\nu'\omega} & T_{r_R}(\iv_{1V},\iv_{2V},\iv_{3V},\iv_{4V}; \iv, \iv', \iw) \\
           &  \times\ \hat U_{l_R}^\FF(\iv) \hat U_{l'_R}^\FF(\iv') \hat U_{m_R}^\Btilde(\iw).
\end{split}
\label{eq:E}
\end{equation}
The tensor network representation of Eq.~(\ref{eq:E}) is given in Fig.~\ref{fig:E}.  Exploiting that
internal structure, one can compute $E G$ and $E^\dagger \hat G$ at a cost of $\bigO(L^5)$ with
a negligible memory overhead.  We discuss this algorithm in Appendix~\ref{app:expand}.

We can now solve Eq.~(\ref{eq:g4matrixfit}) efficiently using a conjugate gradient solver
based on Lanczos bidiagonalization, as these solvers only require us to compute
matrix--vector products  $E G$ and $E^\dagger\hat G$ for given $G$ and $\hat G$ instead of creating the
full $E$. These solvers come
with a number of guarantees, in particular exponential convergence with the
number of matrix--vector products, with the convergence rate depending on how $E$ is conditioned~\cite{Saad:SparseLinear}.
Apart from pathological cases, they also guarantee success after constructing the ``full''
matrix, which implies a worst-case runtime scaling of $\bigO(L^8)$ of the fitting
procedure.  In practice, we choose the LSMR solver~\cite{LSMR} and find that for cutoffs not too low, 
it converges in relatively few iterations, typically around 100.

Although $L$ scales only logarithmically with $\beta\wmax$,
the power in the scaling may become problematic in calculations with large $L$, e.g. at low $T$ with a small cutoff $\epsilon_L$.
We may be able to improve on this scaling using the low-rank approximation of an IR basis vector (tensor network representations)~\cite{Shinaoka:2020ji}.

\begin{figure}
    \centering
    \includegraphics[scale=.65]{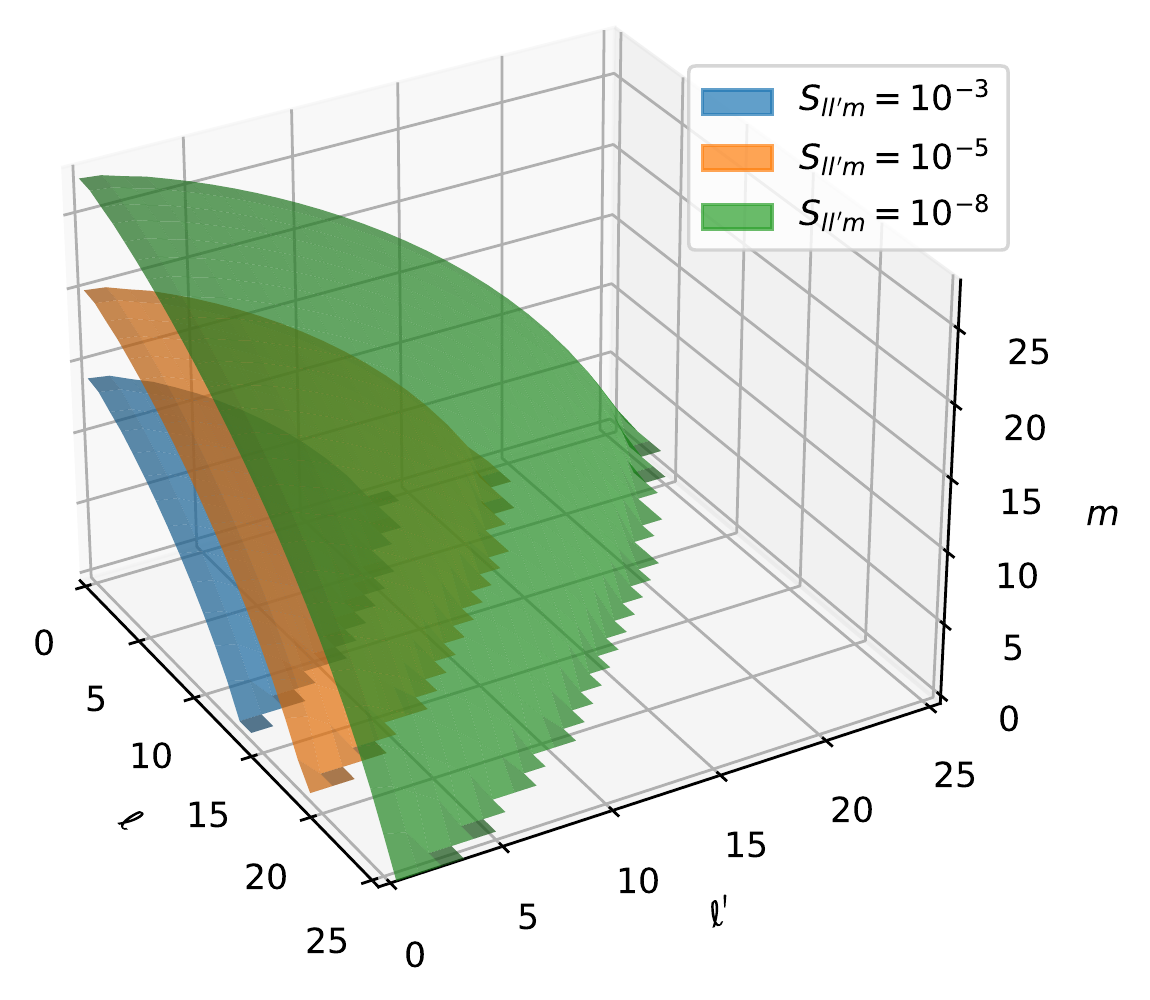}
    \caption{Truncation of singular values $S_{ll'm}=S^\FF_l S^\FF_{l'} S^\BB_m$ in order
             to combat overfitting: smoothened isosurfaces of $S_{ll'm}$ for $\beta\wmax=100$
             for different cutoffs of $10^{-3}$ (innermost), $10^{-5}$, $10^{-8}$ (outermost).
             }
    \label{fig:overfitting}
\end{figure}

Let us mention one complication in solving Eq.~(\ref{eq:g4matrixfit}): examining 
Eq.~(\ref{eq:rho4ir}), we see that $G_{r,ll'm}$ must be decay
as $S_{ll'm}=S^\FF_l S^\FF_{l'} S^\Btilde_m$.  This implies that for a
given cutoff $\epsilon_L\sim S_L/S_0$, we are including terms $\rho_{ll'm}$ that are
dampened below the level of $\epsilon_L$.  We illustrate in Fig.~\ref{fig:overfitting},
where we plot the isosurfaces of $S_{ll'm}$ for different error levels.  Basis
coefficients outside of the isosurface  cannot be reliably fitted by empirical $\hat G$ with
errors at the same level, and including them may thus lead to overfitting. 

One can however remedy this by restricting $\mathcal R$ to:
\begin{equation}
    \mathcal R = \big\{ (r, l, l', m) : S_{ll'm}/S_{000} \ge S^\alpha_L/S^\alpha_0\big\},
\end{equation}
i.e., only to those coefficients which are not dampened below the tolerance.  Since
one-particle singular values $S^\alpha_l$ decay faster than $\exp(-c l)$ but slower
than $\exp(-c l^2)$ for all real coefficients $c$, we have $2L^3 < |\mathcal R| < 6L^3$,
and typically $|\mathcal R|\approx 4L^3$.  In addition to mitigating oversampling,
we have thus also reduced the number of coefficients needed for modeling $G$ by
a factor of three.  Since in practice $|\WW| > 6L^3$, we have also turned
Eq.~(\ref{eq:g4matrixfit}) from a formally underdetermined to an overdetermined system.

% ===============================================
\section{Numerical benchmarks}
\label{sec:bench}
% ===============================================

We now move to benchmark the method on physical examples and provide numerical evidence
for the exponential convergence.  One of the simplest test cases is the Anderson impurity
model.  Its Hamiltonian reads:
\begin{equation}
\begin{split}
H &= \sum_{ab=1}^{\Norb} E_{ab} \cdag_a\cee_b
   + \frac14 \sum_{abcd=1}^{\Norb} U_{abcd} \cdag_a\cdag_b\cee_d\cee_c \\
&+ \sum_{k=1}^{\Nbath} \sum_{a=1}^{\Norb} (V_{ka} \fdag_k \cee_a + V^*_{ka} \cdag_a \eff_k)
   + \sum_{k=1}^{\Nbath} E'_k \fdag_k \eff_k,
\end{split}
\label{eq:aim}
\end{equation}
where $\cee_a$ annihilates an electron on the impurity spin-orbital $a$, $a=1,\ldots,\Norb$,
$\eff_k$ annihilates an electron on the bath spin-orbital $k$, $k=1,\ldots,\Nbath$,
$E_{ab}$ parametrizes the impurity levels, $U_{abcd}$ is the (antisymmetrized) two-body
interaction strength, $V_{ka}$ are the hybridization strengths, and $E'_k$ are the bath level energies.

For $\Norb+\Nbath$ small enough, one can compute the full vertex $F$ (\ref{eq:F}) to arbitrary precision
using exact diagonalization.  However, even with the full vertex exact, only one or two digits of accuracy in
the irreducible vertices $\Gamma$ are achievable with existing methods before the computational
resources are exhausted.

For a convergence analysis across several orders of magnitude, we thus resort to
limiting cases of the Anderson impurity model for which analytical results are
available: the atomic limit (Sec.~\ref{sec:atom}) and the weak-coupling limit
(Sec.~\ref{sec:weak}).

% -----------------------
\subsection{Hubbard atom}
\label{sec:atom}
% -----------------------

We first consider the Hubbard atom, which is the zero-hybridization limit, $V\to 0$, of the
half-filled single impurity Anderson  model (\ref{eq:aim}).  Its Hamiltonian reads:
\begin{equation}
H = U \cdag_\uparrow \cdag_\downarrow \cee_\downarrow \cee_\uparrow 
- \frac U2 (\cdag_\downarrow \cee_\downarrow + \cdag_\uparrow \cee_\uparrow),
\label{eq:hubbard-atom}
\end{equation}
where $\cee_\sigma$ annihilates an electron of spin $\sigma$ and $U$ is the
strength of the electron-electron interaction.  The spectral function
is given by $\rho(\omega) = \frac12[\delta(\omega+\frac U2) + \delta(\omega - \frac U2)]$.

Despite the simplicity of Eq.~(\ref{eq:hubbard-atom}), the irreducible and reducible
vertices of the Hubbard atom have highly nontrivial structures in the Matsubara frequency
domain:  there are sharp, ``$\delta$-like'' planes running horizontally, vertically and
diagonally through the frequency box, structures which do not decay
asymptotically.  Moreover, the proximity of a family of poles on the imaginary
frequency axis~\cite{Schaefer:PRL13,Schaefer:PRB16} as well as channel coupling of the dominant spin susceptibility and
the exponentially suppressed charge susceptibility\cite{Krien:PRB19} causes the irreducible vertex
to vary across several order of magnitudes.  As a result, solving the BSE for the
atomic limit presents a formidable challenge to solvers which truncate the
Matsubara frequency domain to a finite frequency box.

\begin{figure}
    \centering
    \includegraphics[width=\columnwidth]{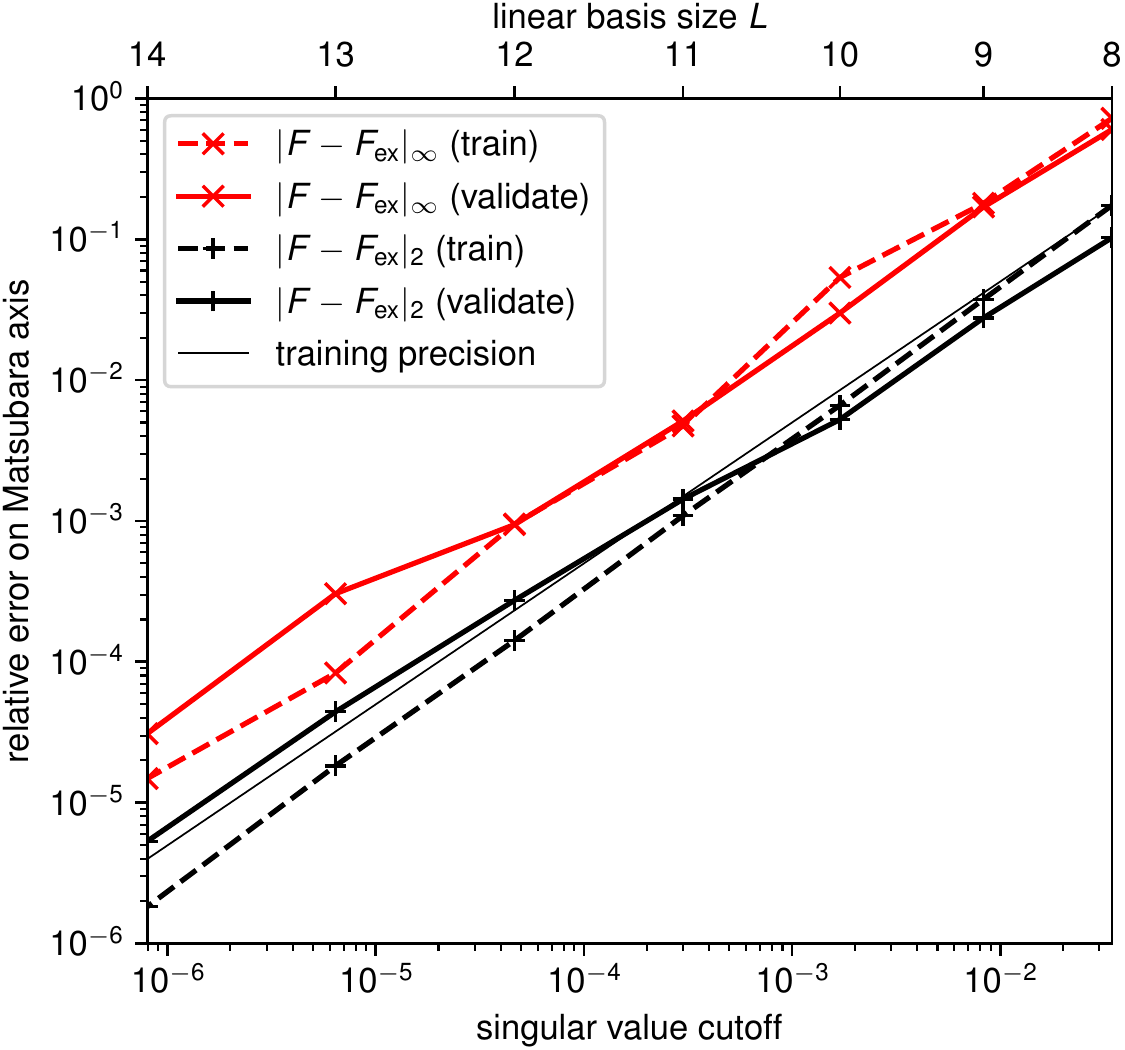}
    \caption{
    Sparse modeling of the Bethe--Salpeter equation for the Hubbard atom with $U = 2.3$,
    $\beta = 1.55$.   We plot the error of the irreducible vertex $F$ on the
    Matsubara axis (normalized by the largest value of $F$) over the singular
    value cutoff (bottom axis) or, equivalently, the IR basis cutoff $L$ (top axis).
    Dotted lines mark the training error (error on the sampling frequencies $\WW'$),
    solid lines indicate the validation error, computed on a frequency box $\WW''$ (\ref{eq:wvalidate4}).
    Black plusses denote the least-squares deviation, while red crosses mark the maximum
    deviation from the analytical result.
    }
    \label{fig:atom_train_test}
\end{figure}

Fortunately, in Ref.~\onlinecite{Thunstroem:PRB18} analytical results are derived for $F$ and
$\Gamma$ for the Hubbard atom.  We are thus able to perform an absolute convergence
analysis of our sparse modeling approach to solving the BSE: we construct
$A_{\Gamma}$ in Eq.~(\ref{eq:AF}) using the analytical expressions for $\Gamma$, use
Eq.~(\ref{eq:bseir}) to solve the BSE, and finally compare the resulting $F$
to the analytical expression $F_\mathrm{ex}$. 
In the following we (arbitrarily) choose $\beta = 1.55$, $U = 2.3$~\footnote{We show the results for the density vertex. For this choice of parameters the only energy scale of the Hubbard atom is $T/U=0.28$ and lies slightly above the first divergence line at $T/U=\sqrt{3}/(2\pi)\approx0.276$~\cite{Schaefer:PRB16}} (for results at other values of $\beta$ see Appendix~\ref{app:divergence}).
We use the IR basis for $\beta\wmax = 10$.

Figure~\ref{fig:atom_train_test} shows the fitting error in $F$ for different
choices of IR basis truncation $L$, which corresponds to different cutoffs $\epsilon$ for
the singular values.  The fitting was performed using the LSMR iterative solver 
(cf.~Sec.~\ref{sec:expansion}) and the system was regularized by imposing an 
accuracy goal of $\norm{F-F_\mathrm{ex}}_2 = 5\epsilon \norm{F}_2$
(black thin line).

The black plusses indicate the least squares deviation
 $\norm{F-F_\mathrm{ex}}_2$, while red crosses indicate the maximum
deviation  $\norm{F-F_\mathrm{ex}}_\infty$.  Both values are normalized by
 $\norm{F}_\infty$, the largest value of $F$.  We see that the ``training'' error  indicated by dotted lines,
i.e., the error on the sampling frequencies $\WW'$, closely tracks
the desired singular value cutoff.  This shows that the IR representation
(\ref{eq:g4ir}) is able to fully capture features of the vertex across multiple
orders of magnitude without any underfitting.

To check our fitting we construct a set of validation frequencies:
\begin{equation}
\begin{split}
    \WW'' = \big\{ (\iw, \iv, \iv') : &\ \nu,\nu'\in \{-29\tfrac\pi\beta,\ldots,29\tfrac\pi\beta\}; \\
                                      &\ \omega\in \{-28\tfrac\pi\beta,\ldots,28\tfrac\pi\beta\}
     \big\} \setminus \WW',
\end{split}
\label{eq:wvalidate4}
\end{equation}
i.e., a dense frequency box of 30 fermionic and 29 bosonic frequencies centered around the origin,
with the sampling frequencies removed.  Fig.~\ref{fig:atom_train_test} shows that the validation
error (solid lines) follows the training error (dotted lines) closely, both for the maximum and
average deviation, which implies there is no significant overfitting and the basis has
predictive power at the accuracy level specified by the training (for results obtained for data with statistical noise see Appendix~\ref{app:noise}).

Let us finally direct our attention to the error scaling with the truncation $L$ of the IR basis,
shown as top axis.  We see that by doubling $L$, we gain more than four orders of magnitude in
terms of precision.

Next, we compare the error scaling of sparse modeling with the conventional (dense) approach:
in the latter, one constructs the operator $A_{\Gamma}$ (\ref{eq:AF}) on a box of $N\times N$
fermionic frequencies centered around the origin and then solves Eq.~(\ref{eq:bse3}) by matrix inversion.
Fig.~\ref{fig:teaser}, shown in the Introduction, compares the validation error on $\WW''$ of sparse modeling with cutoff $L$ and of
the conventional approach with box size $N$.  Let us note that this is not a fair comparison in
terms of computational time, rather, the point is the \emph{scaling} of the error: we see that the
error improves as $1/N$, which together with the factor that one requires for storage of $N^2$ frequencies,
makes it difficult to add precision once $N$ becomes sufficiently large. On the other hand,
sparse modeling converges quickly with cutoff.

% ------------------------------
\subsection{Multi-orbital weak coupling}
\label{sec:weak}
% ------------------------------

We will now consider the opposite limit of the Anderson impurity model (\ref{eq:aim}):
the limit of weak coupling. There, it is reasonable to approximate the irreducible
vertex with the bare vertex $U$:
\begin{equation}
    \hat\Gamma_{abcd}(\iw,\iv,\iv') \approx U_{abcd},
    \label{eq:gammawc}
\end{equation}
where we now consider more than one orbital and thus $\Gamma$ acquires spinorbital indices
$a,b,c,d$.
Similarly, the one-particle Green's function can be approximated by its non-interacting
counterpart:
\begin{equation}
\hat G^{-1}_{ab}(\iv) \approx \hat G^{-1}_{0,ab}(\iv) = \iv\delta_{ab} - E_{ab} - 
\underbrace{\sum_{k=1}^{\Nbath} \frac{V_{ka}^* V_{kb}}{\iv - E'_k}}_{\Delta_{ab}(\iv)},
    \label{eq:poles}
\end{equation}
where $E$, $E'$ and $V$ are defined in Eq.~(\ref{eq:aim}) and we have introduced the
hybridization function $\Delta(\iv)$.

The Bethe--Salpeter equation (\ref{eq:bse3}) in this multi-orbital case then takes the
form:
\begin{equation}
\begin{split}
    &\hat\Gamma_{abgh}(\iw;\iv,\iv'') + \frac{1}{\beta}\sum_{\nu'} \sum_{cdef} \hat\Gamma_{abcd}(\iw;\iv,\iv')\hat G_{de}(\iv') \\
    &\quad\times\hat G_{fc}(\iv'+\iw) \hat F_{efgh}(\iw;\iv',\iv'') = \hat F_{abgh}(\iw;\iv,\iv'').
\end{split}
\label{eq:bse-multi}
\end{equation}

By iterating Eq.~(\ref{eq:bse-multi}) with $F_0 = \Gamma = U$ and using the residual
calculus (cf.~Appendix~\ref{app:residual}), one can show that $F$ is given
as the solution to the following algebraic equation:
\begin{equation}
    U_{abgh} + \sum_{cdef} U_{abcd} \hat\chi_{0,cdef}(\iw) \hat F_{efgh}(\iw) = \hat F_{abgh}(\iw),
    \label{eq:bseweak}
\end{equation}
where we have introduced:
\begin{equation}
    \hat\chi_{0,cdef}(\iw) = \sum_{ij} \frac{f(\gamma_i)-f(\gamma_j)}{\iw+\gamma_i-\gamma_j} g_{id}^* g_{ie} g_{jf}^* g_{jc}
    \label{eq:chi0}
\end{equation}
and $f(x)$ is the Fermi function, and $\gamma_i$ and $g_j$ are the eigenvalues and eigenvectors,
respectively, of the one-body matrix formed blockwise by $(E, V, V^\dagger, E')$.
(For $\gamma_i=\gamma_j$, the corresponding term in Eq.~(\ref{eq:chi0}) has to be understood in the limit $\gamma_i\to \gamma_j$.) 

\begin{figure}
    \centering
    \includegraphics[width=\columnwidth]{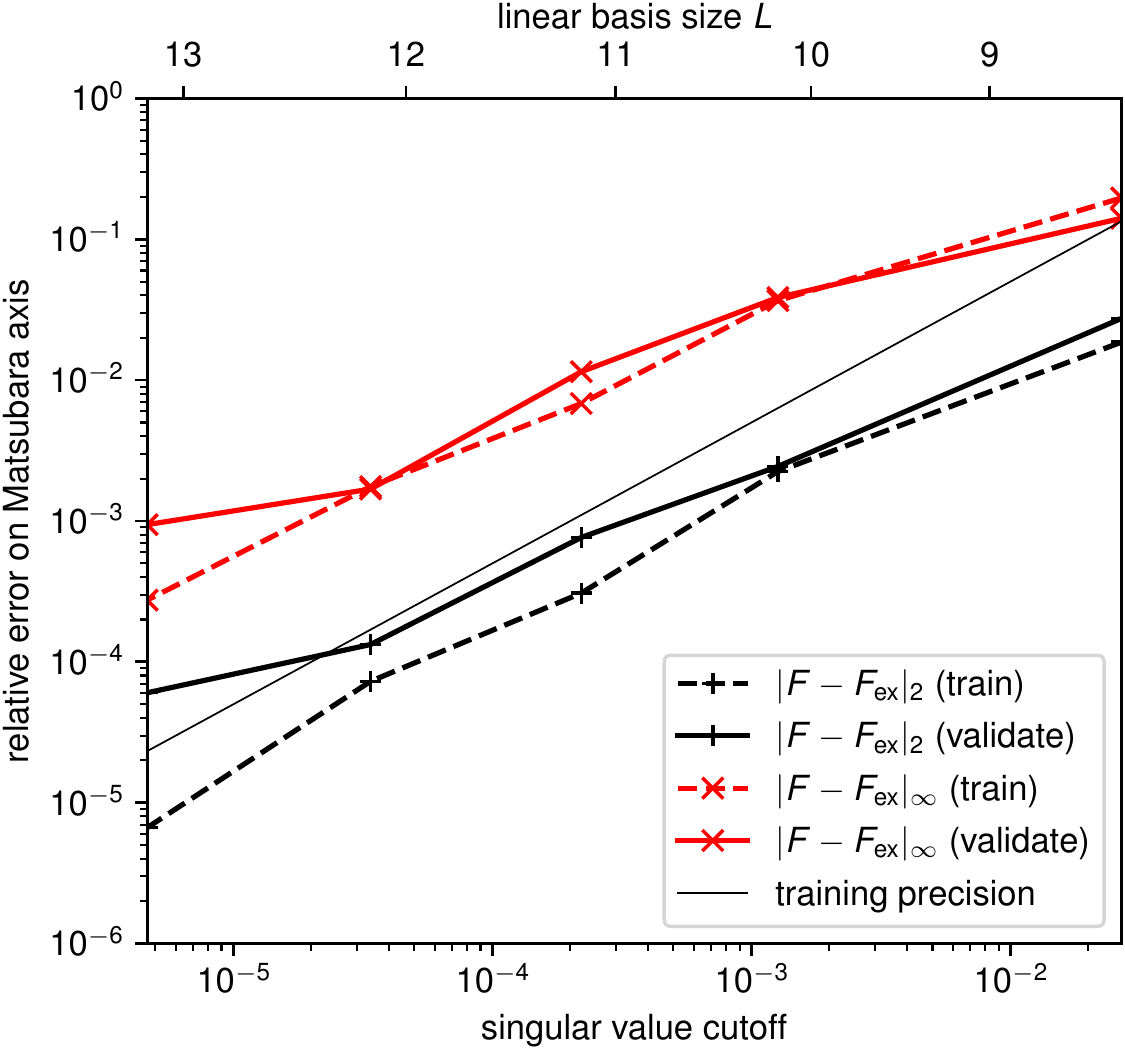}
    \caption{
    Sparse modeling of the Bethe--Salpeter equation for the multi-orbital weak-coupling limit with $U = 0.3$, $\beta = 1.55$ ($\beta\wmax=10$).
    We plot the error of the full vertex $F$ on the
    Matsubara axis (normalized by the largest value of $F$) over the singular
    value cutoff (bottom axis) or, equivalently, the IR basis cutoff $L$ (top axis).
    Dotted lines mark the training error (error on the sampling frequencies $\WW'$),
    solid lines indicate the validation error, computed on a frequency box $\WW''$ (\ref{eq:wvalidate4}).
    Black plusses denote the least-squares deviation, while red crosses mark the maximum
    deviation from the analytical result.
    }
    \label{fig:multi_orbital_train_test}
\end{figure}

We note that in this approximation $F$ has no dependence on fermionic frequencies. 
By combining a pair of spinorbital indices into a single index, Eq.~(\ref{eq:bseweak}) can be
transformed into a system of linear equations and solved exactly.  It thus provides an
ideal benchmark for solving Eq.~(\ref{eq:bse-multi}) with our sparse modeling
approach.

Fig.~\ref{fig:multi_orbital_train_test} shows the fitting error in $F$ for different
choices of IR basis truncation $L$, which corresponds to different cutoffs $\epsilon$ for
the singular values.
We considered three impurity spinorbitals ($\Norb=3$) and four bath sites ($\Nbath=4$): $U_{abba}=0.3$ ($a\neq b$) or 0 (otherwise), $\qty{\gamma_i}=\qty{-0.25, -0.1, 0.1, 0.25}$, $g_{ia}=\cos(i+3a/2+1/10)$ for $a=1,2,3$ and $i=1,2,3,4$.
We used $\beta=1.55$.
Similarly to Fig.~\ref{fig:atom_train_test},
Fig.~\ref{fig:multi_orbital_train_test} shows the least squares deviation
$\norm{F-F_\mathrm{ex}}_2$ and the maximum deviation $\norm{F-F_\mathrm{ex}}_\infty$ for the same validation frequencies. 
One can see that there is neither significant overfitting nor underfitting. 
The result shows that the present method works for multi-orbital systems.

% ===============================================
\section{Results for the Anderson impurity model}
\label{sec:impurity}
% ===============================================

\begin{figure}
    \centering
    \includegraphics[width=\columnwidth]{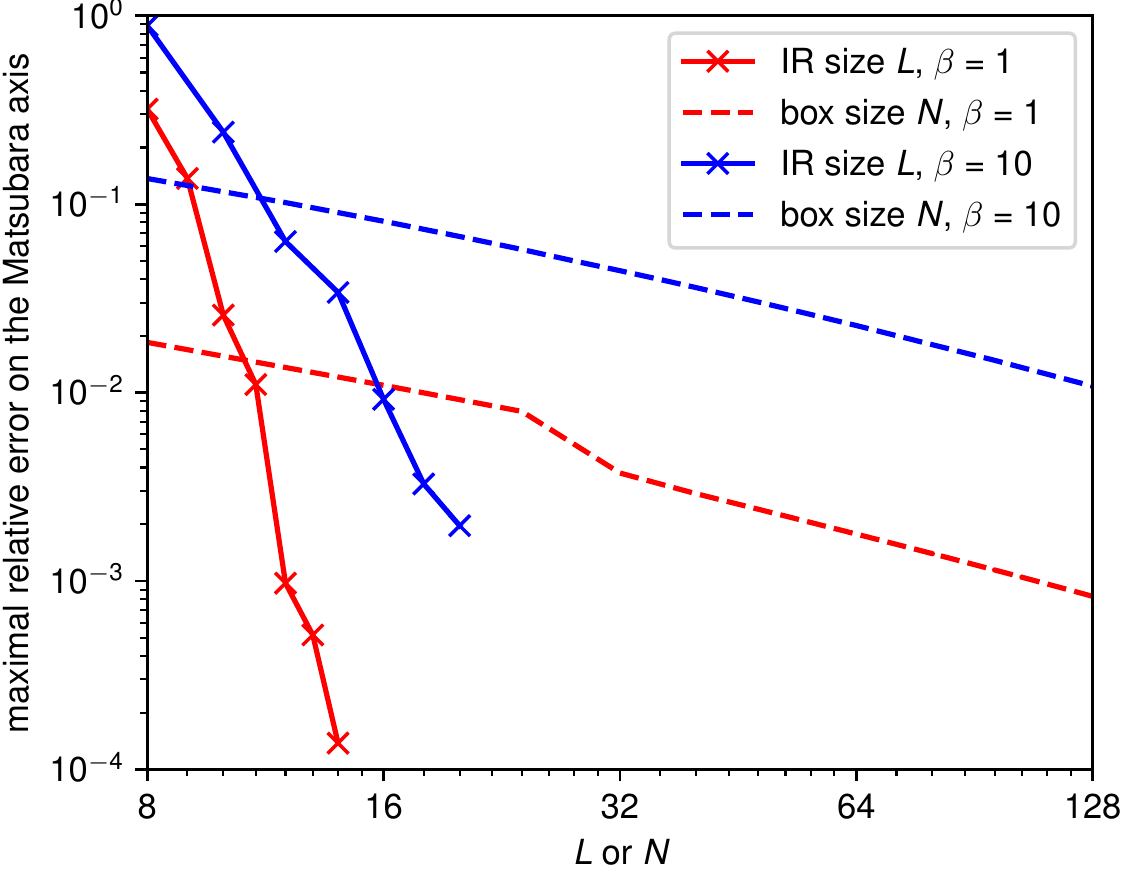}
    \caption{
    Comparison between the maximal relative deviation $\norm{F-F_{ex}}_\infty/\norm{F}_\infty$ of sparse modeling of the Bethe--Salpeter equation with the dense (box) calculation for the impurity model with constant hybridization function $\Delta=\pi/5$ for  $U/\Delta = 1.59$ and two values of the inverse temperature $\beta = 1$ ($\beta\wmax=10$) and $\beta= 10$ ($\beta\wmax=100$).
    }
    \label{fig:impurity_test}
\end{figure}

In this section we apply the method to solve the BSE for a fully-fledged Anderson impurity model \eqref{eq:aim}, where analytical expressions for the vertices are not known. As already mentioned in Sec.~\ref{sec:bench}, although this model can be solved exactly, it is not possible to construct benchmark two-particle vertex functions with arbitrary accuracy. The vertices can be however obtained numerically with the precision of several digits from the parquet equations method~\cite{Bickers04,victory}. 

In the following we will also use the parquet approximation (PA). PA is not exact but gives excellent results in the weak coupling regime~\cite{Hille2020,Schaefer2020}. It has the advantage, that the two-particle vertices do not have statistical errors. Contrary to the weak-coupling model used for benchmarking in Sec.~\ref{sec:bench}, both the irreducible vertex $\Gamma$ and the full vertex $F$ are dependent on two fermionic and one bosonic frequency and have nontrivial structure coming from channel mixing in the parquet equations. 

The Anderson impurity model is characterized by:
(i) The strength of impurity-bath hybridization function $\Delta_{ab}(\nu)$ [cf.~Eq.~(\ref{eq:poles})].  We choose it to be spinorbital and energy independent with $\Delta=\pi/5$~\footnote{The same hybridization function was used in Refs.~\cite{Chalupa2018, Chalupa2020}}. We restrict ourselves to two spinorbitals (spins) on the impurity and two spins in the bath.
(ii) The interaction $U$ on the impurity between electrons with different spins, which is here $U/\Delta=1.59$,  corresponding to weak coupling.
(iii) The impurity filling $n$ (here $n=1$, i.e. half-filling).
For these parameters the estimated Kondo temperature is $T_K\approx0.36$ but due to small value of $U/\Delta$ we are far from vertex divergencies present in this model~\cite{Chalupa2018,Chalupa2020}. We consider two temperatures $T=1$ and $T=0.1$.  

In Fig.~\ref{fig:impurity_test} we show a comparison between maximal relative deviation $\norm{F-F_{ex}}_\infty/\norm{F}_\infty$  of dense (with linear size of the frequency box $N$) and sparse (with IR basis size $L$) evaluations of the BSE for two different inverse temperatures (red: $\beta=1$, blue: $\beta=10$). The input irreducible vertex  $\Gamma$ and the benchmark vertex $F_{ex}$ were obtained from a PA solution on a frequency box with linear size $N=1024$ \footnote{Concretely, we show the results for the magnetic channel.}. The precision of this benchmark calculation, as estimated from box-size convergence, is $10^{-5}$ for $\beta=1$ and $10^{-3}$ for $\beta=10$, which limits our comparison to only a couple of orders of magnitude for the lower temperature. It is however already visible, that also in this case the error drops quickly with $L$ as compared to the $\bigO(1/N)$ scaling of the dense calculation. We show here only the larger, maximal relative deviation on the validation set $\WW''$~\eqref{eq:wvalidate4}. The average relative deviation $\norm{F-F_{ex}}_2$ is, similarly to the atomic limit, smaller and has the same scaling behaviour as the maximal one. 

Comparing the results for the two different inverse temperatures, we see that reaching the same precision for an order of magnitude lower temperature requires only twice the IR basis size $L$, whereas $N$ needs to be at least an order of magnitude larger.

% ===============================================
\section{Conclusions and outlook}
\label{sec:conclusion}
% ===============================================
We proposed an efficient method for solving the Bethe--Salpeter equation based on sparse modeling and the intermediate representation (IR).
Our algorithm is based on a sparse convolution method, which allows us to perform
summation over frequencies of inner propagators needed in solving the Bethe--Salpeter equation.
All intermediate objects, such as vertices, are stored in compressed form.
We numerically demonstrated the exponential convergence of the algorithm with respect to the basis size for the Hubbard atom, the weak-coupling limit of a multi-orbital impurity, and a realistic impurity problem. In the present study, we focused on the particle-hole channel, however,
the proposed method can be straightforwardly applied to the particle-particle channel as well.

In Ref.~\onlinecite{Shinaoka:2020ji}, some of the authors and co-workers introduced a tensor network representation of the two-particle Green's functions, which allows us to compress further  multidimensional data with many indices for the IR basis: spins and orbitals. Combining the present method and the tensor network representation will open a new avenue to efficient calculation of two-particle response functions of correlated materials in e.g. DFT(\textit{GW})+BSE~\cite{Blase2020,OnidaRMP,Aryasetiawan_1998,Takada2001,Maggio2017,Pavlyukh2020}, DFT+dynamical mean-field theory~\cite{GeorgesRMP,Kunes11,Rohringer2012, Hafermann2014} and non-local extensions thereof~\cite{Lin2012,Otsuki2014,RohringerRMP},
 or functional renormalization group~\cite{Metzner2012,dmf2rg,Kugler2018}. 

The natural connection of the IR basis to the analytic continuation kernel charts a course to
obtaining two-particle response functions on the real-frequency axis from Matsubara data, though we expect challenges of regularization and bias similar to the one-particle case will have to be overcome first.  Ultimately, this may allow the interpretation of experimental spectroscopy, optical conductivity and neutron scattering data for models and parameter regimes where direct calculation in real frequencies is infeasible.
 
Also diagrammatic calculations based on more numerically involving equations, such as parquet equations~\cite{DeDominicis,Vasiliev98,Bickers04}, and calculations involving higher-order vertices can be made possible by combining the IR basis for frequencies with the form-factor basis for momenta~\cite{Eckhardt2020, Eckhardt2018}. 

It was recently shown that irreducible vertices diverge on specific lines in the parameter space~\cite{Schaefer:PRL13,Schaefer:PRB16}.
This causes numerical instability in solving diagrammatic equations near the divergence line.
Such divergence is characterized by emergent poles on an imaginary frequency axis, which the original IR basis does not take into account.
Further augmentation of the IR basis may provide a controllable way to numerically handle effects of the vertex divergence.

The unit-tested implementation, with which the data in this manuscript
has been generated, is available from the authors upon request.  We expect to release it as an
open-source package in the near future.

% -----------------------------------------------

\begin{acknowledgments}
We are indebted to F. Krien, K. Held, S. Huber,  E. Gull, J. Kune\v{s}, T. Nomoto, E. Maggio, and J. Tomczak for fruitful discussions and 
careful review of the manuscript.
H.S. was supported by JSPS KAKENHI Grants No.~18H01158, No.~21H01003, and No.~21H01041.  M.W. and A.K. acknowledge support from the Austrian Science Funds (FWF) through project P30997.
The calculations were done in part on the Vienna Scientific Cluster (VSC).
We used the irbasis library~\cite{irbasis} for computing IR basis functions.
\end{acknowledgments}

%% =========================== APPENDIX ===========================

\appendix

% -----------------------
\section{Residual calculus}
\label{app:residual}
% -----------------------

For completeness, this appendix is deriving Eq.~(\ref{eq:absplit}) from Eq.~(\ref{eq:prod}).
We begin by restating Eq.~(\ref{eq:prod}): let
$A$ and $B$ be one-particle Green's functions and let $C$
be defined as follows:
\begin{equation}
C(\iw)=\frac1\beta\sum_{\nu}A(\iv)B(\iv+\iw).
\label{eq:prod2}
\end{equation}
We expand $A$ and $B$, respectively,
into a set of poles $a_{i}$ and $b_{i}$ with expansion coefficients
$A_{i}$ and $B_{i}$: 
\begin{equation}
C(\iw)=\frac1\beta
\sum_{\nu}\sum_{i,j}\frac{A_i B_j}{(\iv-a_i)(\iv+\iw-b_j)}.
\label{eq:2poles}
\end{equation}
The sum (\ref{eq:2poles})
can be performed explicitly via residual calculus, yielding the Lindhard bubble:
\begin{equation}
  C(\iw) = \sum_{i,j} \frac{A_i B_j}{\iw + a_i - b_j} \big(f(a_i) - f(b_j + \iw)\big),
  \label{eq:residual}
\end{equation}
where $f(x)$ is the Fermi function. We note that for a bosonic Matsubara frequency,
$f(x+\iw)=f(x)$, while for a fermionic Matsubara frequency, $f(x+\iw)=b(x)$, where
$b(x)$ is the Bose function.  By expanding $f(x)$ in its Matsubara sum, we
find:
\begin{equation}
  C(\iw) = \frac 1\beta \sum_\nu \sum_{i,j} \frac{A_i B_j}{\iw + a_i - b_j} \Big[ 
  \frac{\ee^{\iv0^-}}{\iv - a_i} - \frac{\ee^{\iv0^-}}{\iv + \iw - b_j} \Big].
  \label{eq:residual-expand}
\end{equation}

Let us briefly comment on the inclusion of Eq.~(\ref{eq:residual}): this may seem like a detour,
as a partial fraction decomposition of Eq.~(\ref{eq:2poles}) yields Eq.~(\ref{eq:residual-expand}) 
for each $\iv$. However, splitting the terms into two sums, there is an ambiguity
in the convergence factor $\exp(\iv0^\pm)$.  This ambiguity must be spurious as the series
(\ref{eq:prod2}) is convergent, yet a convergence factor of $\exp(\iv0^+)$ will give
an overall sign in Eq.~(\ref{eq:residual}).  The proper way to split up this sum is
using the residual calculus, which fixes the convergence factor to be $\exp(\iv0^-)$.

Finally, comparing coefficients in Eqs.~(\ref{eq:prod2}) and (\ref{eq:residual-expand}) yields
Eq.~(\ref{eq:absplit}), which we restate here for convenience:
\begin{equation}
  C(\iw) = A'_\omega(\iv) + B'_\omega(\iv+\iw),
  \label{eq:absplit2}
\end{equation}
where $A'$ and $B'$ are now auxiliary Green's functions.

\begin{figure}
% -- definitions
\SetKwInput{KwInput}{Input}
\SetKwInput{KwCommonInput}{Common input}
\SetKwProg{KwFunction}{Function}{ is}{end}
\SetKwFunction{FApply}{apply}
\SetKwFunction{FAdjoint}{adjoint}
% -- the actual algorithm
\begin{algorithm}[H]
    \KwCommonInput{
    \begin{itemize}[leftmargin=*,label={--},noitemsep]
        \item Tuple $I = ((\nu_1, \nu'_1, \omega_1), \ldots, (\nu_N, \nu'_N, \omega_N))_{n=1,\ldots,N}$,
        where $\iv$ and $\iv'$ are fermionic frequencies, $\iw$ is a bosonic
        frequency, and the elements of $I$ are in strict 
        lexicographical order. $I_0:=(-\infty,0,0)$.
    \end{itemize}
    }
    \KwFunction{\FApply{$\rho$}}{
        \KwInput{$\rho_{ll'm}$, a $L\times L\times L$ complex tensor}
        \KwResult{$G_n$ through Eq.~(\ref{eq:Ecore})}
        \KwData{$A$, a $L\times L$ tensor, and $B$, and $L$ vector}
        \For{$n=1,\ldots,N$}{
            \If{$\nu_n \ne \nu_{n-1}$}{
                $A_{l'm} \gets \sum_{l=0}^{L-1} U^\FF_l(\iv_n) \rho_{ll'm}$  \\
            }
            \If{$\nu_n \ne \nu_{n-1}\vee \nu'_n \ne \nu'_{n-1}$}{
                $B_m \gets \sum_{l'=0}^{L-1} U^\FF_{l'}(\iv'_n) A_{l'm} $\\
            }
            $G_n \gets \sum_{m=0}^{L-1} U^\Btilde_m(\iw_n) B_m$\\
        }  % end for
    }  % end function
    \KwFunction{\FAdjoint{$G$}}{
        As in apply, but with matrices replaced by their adjoints and steps
        done in reverse order.
    }  % end function
    \end{algorithm}
    \caption{Algorithm for fast-on-the fly expansion of $E\rho$ using \texttt{apply} and 
    $E^\dagger G$ using \texttt{adjoint}, each in $\bigO(L^5)$ time with $\bigO(L^2)$ auxiliary
    memory.}
    \label{fig:algo}
\end{figure}

% -----------------------
\section{Fast on-the-fly expansion}
\label{app:expand}
% -----------------------

In order to solve the fitting problem (\ref{eq:g4matrixfit}) using a sparse least squares solver,
we have to apply $E$, defined in Eq.~(\ref{eq:E}) to an arbitrary IR basis vector as well as $E^\dagger$
to a sampling frequency vector in an efficient manner.

The core part of applying $E$ is the construction of the following intermediate object in each
channel:
\begin{equation}
    G_n = \sum_{l=0}^{L-1} U^\FF_l(\iv_n) \sum_{l'=0}^{L-1}U^\FF_{l'}(\iv'_n)\sum_{m=0}^{L-1}U^\Btilde_m(\iw_n)\rho_{ll'm},
    \label{eq:Ecore}
\end{equation}
i.e., the application of the transformation matrices
$\hat U$ followed by the projection to those frequencies which, after translation using $T_r$, will
end up in the sampling frequency set.

We note that this structure (cf.~Fig.~\ref{fig:E}), is in principle well-suited
for on-the-fly application, as the $\hat U$ tensors can be applied separately one after the other
and we then simply select elements. The problem is that
the internal frequencies $\iv, \iv', \iw$  not only contain the $L$ one-particle sampling frequencies
in $\WW^\alpha$, but also any shifts of these frequencies due to $T_r$.  In total one has about $4L^2$
unique frequencies for $\iv$ and similarly for $\iv'$ and $\iw$.  Evaluating Eq.~(\ref{eq:Ecore}) from
right to left, we construct an intermediate tensor of size $\bigO(L^6)$ before selecting
$\bigO(L^3)$ elements from it.  This puts the total cost at $\bigO(L^9)$ time.

We can improve upon this by discarding the block structure and simply compute $G_n$
for each $n$ separately.  This involves contracting $\rho_{ll'm}$ along each axis with
three vectors $U^\FF_l(\iv_n)$, $U^\FF_{l'}(\iv'_n)$, and $U^\Btilde_m(\iw_n)$ at a cost of 
$\bigO(L^3)$, $\bigO(L^2)$, and $\bigO(L)$, respectively. Since this
has to be done for each sampling frequency, the total cost is $\bigO(L^6)$.

There is still room for improvement: if we order the sampling frequencies lexicographically,
we can reuse partial contraction results from one sampling point to the next.  Since one
observes only $\bigO(L^2)$ unique $\iv_n$, this brings down the total cost to $\bigO(L^5)$,
while incurring a memory overhead of $\bigO(L^2)$ for storing the partial results.  We
list as function \texttt{apply} in the algorithm in Fig.~\ref{fig:algo}.

The core of the reverse direction, i.e., the application $E^\dagger$ to a sampling frequency vector,
is the construction of the following object in each channel:
\begin{equation}
    \rho_{ll'm} = \sum_n \big[ U^\FF_l(\iv_n) U^\FF_{l'}(\iv'_n) U^\Btilde_m(\iw_n) \big]^* G_n,
    \label{eq:Eadj}
\end{equation}
where $U^*$ denotes the complex conjugate.  A similar idea applies to Eq.~(\ref{eq:Eadj})
as to Eq.~(\ref{eq:Ecore}): we perform the
outer product for a sequence of vectors and cache the intermediate results from frequency
to frequency.  This again yields a $\bigO(L^5)$ cost and $\bigO(L^2)$ auxiliary memory
requirement.

\begin{figure}
    \centering
    \includegraphics[width=\columnwidth]{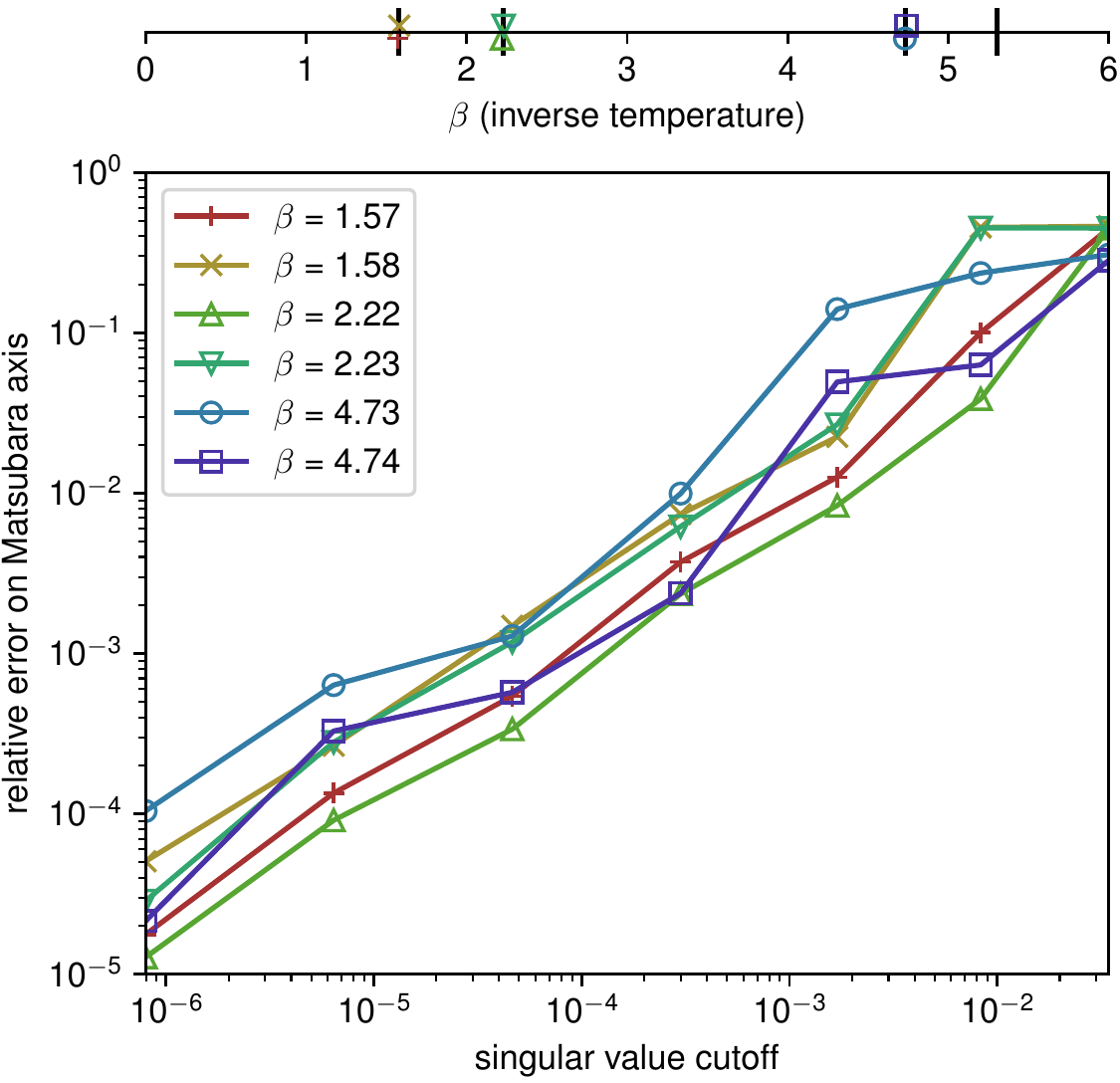}
    \caption{
    Relative error $\norm{F-F_{ex}}_2/\norm{F}_\infty$ on the validation set of Matsubara frequencies for sparse modeling of the Bethe--Salpeter equation in the density channel for the Hubbard atom with $U = 2.3$,
    and different values of $\beta$ in the vicinity of the divergencies of $\Gamma$ that occur for  $U = 2.3$ at $\beta=\{1.577, 2.229, 4.732, \ldots\}$ as shown on the inverse temperature axis above the plot.
    The rest of the calculation  conditions are the same as in Fig.~\ref{fig:atom_train_test}.
    }
    \label{fig:divergencies}
\end{figure}

% -----------------------
\section{Divergences of the irreducible vertex}
\label{app:divergence}
% -----------------------

\begin{figure*}
    \centering
    \includegraphics[width=\textwidth]{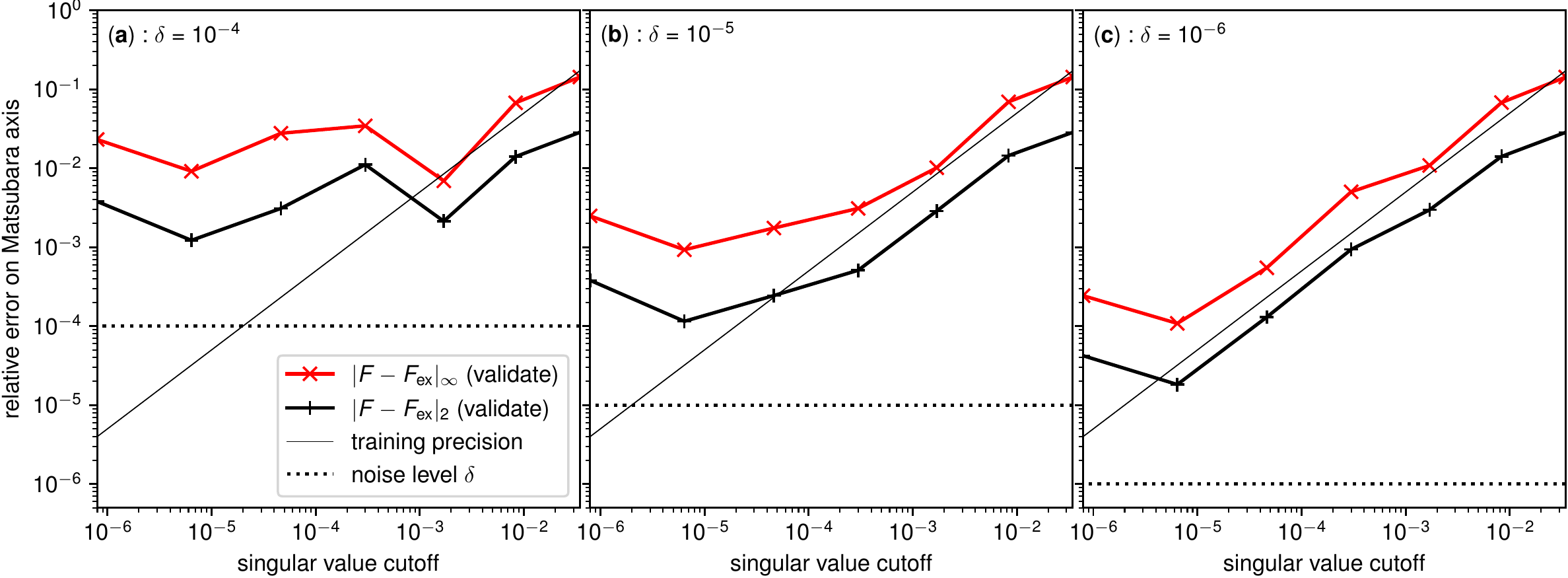}
            \caption{
    Sparse modeling of the Bethe--Salpeter equation for the Hubbard atom with $U = 2.3$ and $\beta = 1.8$ with random Gaussian noise in the input data $\Gamma$.
    The noise level $\delta$ was set to $\delta=10^{-4}$, $10^{-5}$ and $10^{-6}$ for panels a, b, and c, respectively.
    The rest of the calculation conditions are the same as those in Fig.~\ref{fig:atom_train_test}.
    The horizontal dotted lines denote the noise levels, respectively.
    }
    \label{fig:atom_train_test_noise}
\end{figure*}

In Refs.~\onlinecite{Schaefer:PRL13,Schaefer:PRB16} it has been shown that the irreducible vertex $\Gamma$ diverges for certain values of temperature and interaction. These divergencies are also present in the Hubbard atom for a series of $T/U$ values (the ratio $T/U$ is the only energy scale in the atomic limit). The results presented in Sec.~\ref{sec:atom} were obtained for the irreducible vertex in the density channel for $T/U=0.28$, which is very close to, but slightly above  the first divergence point at $T/U=\frac{\sqrt{3}}{2\pi}\approx0.276$. In Fig.~\ref{fig:divergencies} we show the scaling of the relative average error $\norm{F-F_{ex}}_2/\norm{F}_\infty$ on the validation set of Matsubara frequencies (cf. Fig.~\ref{fig:atom_train_test}) for several values of the inverse temperature $\beta=1/T$ (keeping the value of $U=2.3$ constant), that lie in the vicinity of vertex divergencies~\cite{Schaefer:PRB16}: slightly above and slightly below the first, second, and third divergence point. We observe that the closeness to a divergency does not change the exponential convergence of the sparse solver. It does however influence the overall magnitude of the error.    

% -----------------------
\section{Stability against noise}
\label{app:noise}
% -----------------------

In practical calculations, the propagators and vertex functions may sometimes only be known from a stochastic 
method: e.g., $F(\iw; \iv, \iv')$ can be obtained from continuous-time quantum Monte Carlo \cite{GullRMP}, provided the sign problem is not too strong, while $\Gamma(\iw, \iv, \iv')$ is available from bold diagrammatic Monte Carlo \cite{ProkofevBold}, provided that the diagrammatic series is convergent and converges to a physical solution. Vertices obtained in this way contain significant statistical noise. 

Before we discuss the two-particle case, let us briefly review the effect of noise on fitting one-particle propagators and vertices.  There the fitting problem translates to an ordinary least
squares problem, $\min_G \Vert AG - \hat G\Vert^2$.  If we now add white noise on $\hat G$, $\hat G\to \hat G + \delta \Vert G\Vert$, the relative error on $G$ on the fitting problem is bounded by:
\begin{equation}
    \frac{\Vert\Delta G\Vert}{\Vert G\Vert} \le \kappa[A]\cdot \Vert\delta\Vert + \frac{S_L}{S_0},
    \label{eq:noise}
\end{equation}
where $S_l$ are the singular values and $\kappa[A]$ is the condition number of the fitting matrix $A$.
This means that as we increase the size of the basis $L$, we expect the error to drop like the singular
values until we reach the $\kappa$ times the noise level, at which point the error flattens out.
$\kappa$ is thus a measure of ``noise amplification''. Since $\kappa\sim 1$ for $\Sigma$ and $G$, we do not observe any noise amplification there \cite{Li:2020eu}.

For the two-particle vertices and propagators, the picture is similar, with the complications that
(a) the fitting matrix in Eq.~(\ref{eq:bseir}) is generated from a quadrature rule, which involves another fitting problem~\eqref{eq:wquad-fit} and (b) overcompleteness causes the condition number $\kappa$ to deteriorate.  This makes it considerably more complicated to prove the stability of our method against noise in all possible cases.

In order to investigate the stability of the present method against noise for a challenging benchmark, we construct noisy input by adding Gaussian noise to the exact irreducible vertex of the Hubbard atom. 
We add noise to the atomic $\Gamma$ as 
\begin{align}
\Gamma(\iw;\iv,\iv') \to 
\Gamma(\iw;\iv,\iv') + \delta \cdot r(\iw;\iv,\iv') \Vert\Gamma\Vert_\infty,
\end{align}
where $r$ are independent identically distributed Gaussian random variables of mean zero and unit variance, and $\delta~(>0)$ denotes the level of the noise.  

Figure~\ref{fig:atom_train_test_noise} shows the computed results for the sparse modeling of the BSE for the Hubbard atom for $U=2.3$ and $\beta=1.8$ and different noise levels.
The rest of the parameters in the calculations are the same as those in Fig.~\ref{fig:atom_train_test}.
One can see that the relative error of the BSE on the Matsubara axis  $\norm{F-F_{ex}}_2/\norm{F}_\infty$ vanishes exponentially until it hits the (amplified) noise level in the input $\Gamma(\iw;\iv,\iv')$, following the form of Eq.~(\ref{eq:noise}).  While the noise is thus somewhat amplified, the method seems to be fundamentally numerically stable.

Since $F$ comes with approximately white noise when generated from, e.g., continuous-time quantum Monte
Carlo in the interaction expansion, auxiliary-field expansion, and also in the hybridization expansion when using symmetric improved estimators \cite{Kaufmann_improved_estimators}, we expect the picture to remain qualitatively similar there.

While thorough performance analysis of solving the BSE for noisy input data is beyond the scope of this paper, the numerical results presented suggest the robustness of the exponential scaling of the error against noise.  

\bibliography{main}

\end{document}